

\catcode `\@=11 

\def\@version{1.3}
\def\@verdate{28.11.1992}


%
%
%
%
%
%

\font\fiverm=cmr5
\font\fivei=cmmi5	\skewchar\fivei='177
\font\fivesy=cmsy5	\skewchar\fivesy='60
\font\fivebf=cmbx5

\font\sevenrm=cmr7
\font\seveni=cmmi7	\skewchar\seveni='177
\font\sevensy=cmsy7	\skewchar\sevensy='60
\font\sevenbf=cmbx7

\font\eightrm=cmr8
\font\eightbf=cmbx8
\font\eightit=cmti8
\font\eighti=cmmi8			\skewchar\eighti='177
\font\eightmib=cmmib10 at 8pt	\skewchar\eightmib='177
\font\eightsy=cmsy8			\skewchar\eightsy='60
\font\eightsyb=cmbsy10 at 8pt	\skewchar\eightsyb='60
\font\eightsl=cmsl8
\font\eighttt=cmtt8			\hyphenchar\eighttt=-1
\font\eightcsc=cmcsc10 at 8pt
\font\eightsf=cmss8

\font\ninerm=cmr9
\font\ninebf=cmbx9
\font\nineit=cmti9
\font\ninei=cmmi9			\skewchar\ninei='177
\font\ninemib=cmmib10 at 9pt	\skewchar\ninemib='177
\font\ninesy=cmsy9			\skewchar\ninesy='60
\font\ninesyb=cmbsy10 at 9pt	\skewchar\ninesyb='60
\font\ninesl=cmsl9
\font\ninett=cmtt9			\hyphenchar\ninett=-1
\font\ninecsc=cmcsc10 at 9pt
\font\ninesf=cmss9

\font\tenrm=cmr10
\font\tenbf=cmbx10
\font\tenit=cmti10
\font\teni=cmmi10		\skewchar\teni='177
\font\tenmib=cmmib10	\skewchar\tenmib='177
\font\tensy=cmsy10		\skewchar\tensy='60
\font\tensyb=cmbsy10	\skewchar\tensyb='60
\font\tenex=cmex10
\font\tensl=cmsl10
\font\tentt=cmtt10		\hyphenchar\tentt=-1
\font\tencsc=cmcsc10
\font\tensf=cmss10

\font\elevenrm=cmr10 scaled \magstephalf
\font\elevenbf=cmbx10 scaled \magstephalf
\font\elevenit=cmti10 scaled \magstephalf
\font\eleveni=cmmi10 scaled \magstephalf	\skewchar\eleveni='177
\font\elevenmib=cmmib10 scaled \magstephalf	\skewchar\elevenmib='177
\font\elevensy=cmsy10 scaled \magstephalf	\skewchar\elevensy='60
\font\elevensyb=cmbsy10 scaled \magstephalf	\skewchar\elevensyb='60
\font\elevensl=cmsl10 scaled \magstephalf
\font\eleventt=cmtt10 scaled \magstephalf	\hyphenchar\eleventt=-1
\font\elevencsc=cmcsc10 scaled \magstephalf
\font\elevensf=cmss10 scaled \magstephalf

\font\fourteenrm=cmr10 scaled \magstep2
\font\fourteenbf=cmbx10 scaled \magstep2
\font\fourteenit=cmti10 scaled \magstep2
\font\fourteeni=cmmi10 scaled \magstep2		\skewchar\fourteeni='177
\font\fourteenmib=cmmib10 scaled \magstep2	\skewchar\fourteenmib='177
\font\fourteensy=cmsy10 scaled \magstep2	\skewchar\fourteensy='60
\font\fourteensyb=cmbsy10 scaled \magstep2	\skewchar\fourteensyb='60
\font\fourteensl=cmsl10 scaled \magstep2
\font\fourteentt=cmtt10 scaled \magstep2	\hyphenchar\fourteentt=-1
\font\fourteencsc=cmcsc10 scaled \magstep2
\font\fourteensf=cmss10 scaled \magstep2

\font\seventeenrm=cmr10 scaled \magstep3
\font\seventeenbf=cmbx10 scaled \magstep3
\font\seventeenit=cmti10 scaled \magstep3
\font\seventeeni=cmmi10 scaled \magstep3	\skewchar\seventeeni='177
\font\seventeenmib=cmmib10 scaled \magstep3	\skewchar\seventeenmib='177
\font\seventeensy=cmsy10 scaled \magstep3	\skewchar\seventeensy='60
\font\seventeensyb=cmbsy10 scaled \magstep3	\skewchar\seventeensyb='60
\font\seventeensl=cmsl10 scaled \magstep3
\font\seventeentt=cmtt10 scaled \magstep3	\hyphenchar\seventeentt=-1
\font\seventeencsc=cmcsc10 scaled \magstep3
\font\seventeensf=cmss10 scaled \magstep3

\def\@typeface{Computer Modern} 

\def\hexnumber@#1{\ifnum#1<10 \number#1\else
 \ifnum#1=10 A\else\ifnum#1=11 B\else\ifnum#1=12 C\else
 \ifnum#1=13 D\else\ifnum#1=14 E\else\ifnum#1=15 F\fi\fi\fi\fi\fi\fi\fi}

\def\mib{\hexnumber@\mibfam}
\def\syb{\hexnumber@\sybfam}

\def\makestrut{%
  \setbox\strutbox=\hbox{%
    \vrule height.7\baselineskip depth.3\baselineskip width 0pt}%
}

\def\bls#1{%
  \normalbaselineskip=#1%
  \normalbaselines%
  \makestrut%
}

%

\newfam\mibfam 
\newfam\sybfam 
\newfam\scfam  
\newfam\sffam  

\def\em{\ifdim\fontdimen1\font>0 \rm\else\it\fi}

\textfont3=\tenex
\scriptfont3=\tenex
\scriptscriptfont3=\tenex

\def\eightpoint{
  \def\rm{\fam0\eightrm}%
  \textfont0=\eightrm \scriptfont0=\sevenrm \scriptscriptfont0=\fiverm%
  \textfont1=\eighti  \scriptfont1=\seveni  \scriptscriptfont1=\fivei%
  \textfont2=\eightsy \scriptfont2=\sevensy \scriptscriptfont2=\fivesy%
  \textfont\itfam=\eightit\def\it{\fam\itfam\eightit}%
  \textfont\bffam=\eightbf%
    \scriptfont\bffam=\sevenbf%
      \scriptscriptfont\bffam=\fivebf%
  \def\bf{\fam\bffam\eightbf}%
  \textfont\slfam=\eightsl\def\sl{\fam\slfam\eightsl}%
  \textfont\ttfam=\eighttt\def\tt{\fam\ttfam\eighttt}%
  \textfont\scfam=\eightcsc\def\sc{\fam\scfam\eightcsc}%
  \textfont\sffam=\eightsf\def\sf{\fam\sffam\eightsf}%
  \textfont\mibfam=\eightmib%
  \textfont\sybfam=\eightsyb%
  \bls{10pt}%
}

\def\ninepoint{
  \def\rm{\fam0\ninerm}%
  \textfont0=\ninerm \scriptfont0=\sevenrm \scriptscriptfont0=\fiverm%
  \textfont1=\ninei  \scriptfont1=\seveni  \scriptscriptfont1=\fivei%
  \textfont2=\ninesy \scriptfont2=\sevensy \scriptscriptfont2=\fivesy%
  \textfont\itfam=\nineit\def\it{\fam\itfam\nineit}%
  \textfont\bffam=\ninebf%
    \scriptfont\bffam=\sevenbf%
      \scriptscriptfont\bffam=\fivebf%
  \def\bf{\fam\bffam\ninebf}%
  \textfont\slfam=\ninesl\def\sl{\fam\slfam\ninesl}%
  \textfont\ttfam=\ninett\def\tt{\fam\ttfam\ninett}%
  \textfont\scfam=\ninecsc\def\sc{\fam\scfam\ninecsc}%
  \textfont\sffam=\ninesf\def\sf{\fam\sffam\ninesf}%
  \textfont\mibfam=\ninemib%
  \textfont\sybfam=\ninesyb%
  \bls{12pt}%
}

\def\tenpoint{
  \def\rm{\fam0\tenrm}%
  \textfont0=\tenrm \scriptfont0=\sevenrm \scriptscriptfont0=\fiverm%
  \textfont1=\teni  \scriptfont1=\seveni  \scriptscriptfont1=\fivei%
  \textfont2=\tensy \scriptfont2=\sevensy \scriptscriptfont2=\fivesy%
  \textfont\itfam=\tenit\def\it{\fam\itfam\tenit}%
  \textfont\bffam=\tenbf%
    \scriptfont\bffam=\sevenbf%
      \scriptscriptfont\bffam=\fivebf%
  \def\bf{\fam\bffam\tenbf}%
  \textfont\slfam=\tensl\def\sl{\fam\slfam\tensl}%
  \textfont\ttfam=\tentt\def\tt{\fam\ttfam\tentt}%
  \textfont\scfam=\tencsc\def\sc{\fam\scfam\tencsc}%
  \textfont\sffam=\tensf\def\sf{\fam\sffam\tensf}%
  \textfont\mibfam=\tenmib%
  \textfont\sybfam=\tensyb%
  \bls{12pt}%
}

\def\elevenpoint{
  \def\rm{\fam0\elevenrm}%
  \textfont0=\elevenrm \scriptfont0=\eightrm \scriptscriptfont0=\fiverm%
  \textfont1=\eleveni  \scriptfont1=\eighti  \scriptscriptfont1=\fivei%
  \textfont2=\elevensy \scriptfont2=\eightsy \scriptscriptfont2=\fivesy%
  \textfont\itfam=\elevenit\def\it{\fam\itfam\elevenit}%
  \textfont\bffam=\elevenbf%
    \scriptfont\bffam=\eightbf%
      \scriptscriptfont\bffam=\fivebf%
  \def\bf{\fam\bffam\elevenbf}%
  \textfont\slfam=\elevensl\def\sl{\fam\slfam\elevensl}%
  \textfont\ttfam=\eleventt\def\tt{\fam\ttfam\eleventt}%
  \textfont\scfam=\elevencsc\def\sc{\fam\scfam\elevencsc}%
  \textfont\sffam=\elevensf\def\sf{\fam\sffam\elevensf}%
  \textfont\mibfam=\elevenmib%
  \textfont\sybfam=\elevensyb%
  \bls{13pt}%
}

\def\fourteenpoint{
  \def\rm{\fam0\fourteenrm}%
  \textfont0\fourteenrm  \scriptfont0\tenrm  \scriptscriptfont0\sevenrm%
  \textfont1\fourteeni   \scriptfont1\teni   \scriptscriptfont1\seveni%
  \textfont2\fourteensy  \scriptfont2\tensy  \scriptscriptfont2\sevensy%
  \textfont\itfam=\fourteenit\def\it{\fam\itfam\fourteenit}%
  \textfont\bffam=\fourteenbf%
    \scriptfont\bffam=\tenbf%
      \scriptscriptfont\bffam=\sevenbf%
  \def\bf{\fam\bffam\fourteenbf}%
  \textfont\slfam=\fourteensl\def\sl{\fam\slfam\fourteensl}%
  \textfont\ttfam=\fourteentt\def\tt{\fam\ttfam\fourteentt}%
  \textfont\scfam=\fourteencsc\def\sc{\fam\scfam\fourteencsc}%
  \textfont\sffam=\fourteensf\def\sf{\fam\sffam\fourteensf}%
  \textfont\mibfam=\fourteenmib%
  \textfont\sybfam=\fourteensyb%
  \bls{17pt}%
}

\def\seventeenpoint{
  \def\rm{\fam0\seventeenrm}%
  \textfont0\seventeenrm  \scriptfont0\elevenrm  \scriptscriptfont0\ninerm%
  \textfont1\seventeeni   \scriptfont1\eleveni   \scriptscriptfont1\ninei%
  \textfont2\seventeensy  \scriptfont2\elevensy  \scriptscriptfont2\ninesy%
  \textfont\itfam=\seventeenit\def\it{\fam\itfam\seventeenit}%
  \textfont\bffam=\seventeenbf%
    \scriptfont\bffam=\elevenbf%
      \scriptscriptfont\bffam=\ninebf%
  \def\bf{\fam\bffam\seventeenbf}%
  \textfont\slfam=\seventeensl\def\sl{\fam\slfam\seventeensl}%
  \textfont\ttfam=\seventeentt\def\tt{\fam\ttfam\seventeentt}%
  \textfont\scfam=\seventeencsc\def\sc{\fam\scfam\seventeencsc}%
  \textfont\sffam=\seventeensf\def\sf{\fam\sffam\seventeensf}%
  \textfont\mibfam=\seventeenmib%
  \textfont\sybfam=\seventeensyb%
  \bls{20pt}%
}

\lineskip=1pt      \normallineskip=\lineskip
\lineskiplimit=0pt \normallineskiplimit=\lineskiplimit




\def\Nulle{0}  
\def\Aue{1}    
\def\Afe{2}    
\def\Ace{3}    
\def\Sue{4}    
\def\Hae{5}    
\def\Hbe{6}    
\def\Hce{7}    
\def\Hde{8}    
\def\Kwe{9}    
\def\Txe{10}   
\def\Lie{11}   
\def\Bbe{12}   


\newdimen\DimenA
\newbox\BoxA

\newcount\LastMac \LastMac=\Nulle
\newcount\HeaderNumber \HeaderNumber=0
\newcount\DefaultHeader \DefaultHeader=\HeaderNumber
\newskip\Indent

\newskip\half      \half=5.5pt plus 1.5pt minus 2.25pt
\newskip\one       \one=11pt plus 3pt minus 5.5pt
\newskip\onehalf   \onehalf=16.5pt plus 5.5pt minus 8.25pt
\newskip\two       \two=22pt plus 5.5pt minus 11pt

\def\Half{\vskip-\lastskip\vskip\half}
\def\One{\vskip-\lastskip\vskip\one}
\def\OneHalf{\vskip-\lastskip\vskip\onehalf}
\def\Two{\vskip-\lastskip\vskip\two}


\def\rTenPT{10pt plus \Feathering}

\def\TenPT{10pt plus \Feathering} 
\def\ElevenPT{11pt plus \Feathering}

\def\Raggedright{
 \rightskip=0pt plus \hsize
}

\def\Fullout{
\rightskip=0pt
}

\def\Hang#1#2{
 \hangindent=#1
 \hangafter=#2
}

\def\EveryMac{
 \Fullout
 \everypar{}
}



\def\title#1{
 \EveryMac
 \LastMac=\Nulle
 \global\HeaderNumber=0
 \global\DefaultHeader=1
 \vbox to 1pc{\vss}
 \seventeenpoint
 \Raggedright
 \noindent \bf #1
}

\def\author#1{
 \EveryMac
 \ifnum\LastMac=\Afe \OneHalf
  \else \Two
 \fi
 \LastMac=\Aue
 \fourteenpoint
 \Raggedright
 \noindent \rm #1\par
 \vskip 3pt\relax
}

\def\affiliation#1{
 \EveryMac
 \LastMac=\Afe
 \eightpoint\bls{\TenPT}
 \Raggedright
 \noindent \it #1\par
}

\def\acceptedline#1{
 \EveryMac
 \Two
 \LastMac=\Ace
 \eightpoint\bls{\TenPT}
 \Raggedright
 \noindent \rm #1
}

\def\abstract{%
 \EveryMac
 \Two
 \LastMac=\Sue
 \everypar{\Hang{11pc}{0}}
 \noindent\ninebf ABSTRACT\par
 \tenpoint\bls{\ElevenPT}
 \Fullout
 \noindent\rm
}

\def\keywords{
 \EveryMac
 \Half
 \LastMac=\Kwe
 \everypar{\Hang{11pc}{0}}
 \tenpoint\bls{\ElevenPT}
 \Fullout
 \noindent\hbox{\bf Key words:\ }
 \rm
}


\def\maketitle{%
  \Two%
  \EndOpening%
  \MakePage%
}


\def\pageoffset#1#2{\hoffset=#1\relax\voffset=#2\relax}


\def\Autonumber{
 \global\AutoNumbertrue  
}

\newif\ifAutoNumber \AutoNumberfalse
\newcount\Sec        
\newcount\SecSec
\newcount\SecSecSec

\Sec=0

\def\:{\let\@sptoken= } \:  
\def\:{\@xifnch} \expandafter\def\: {\futurelet\@tempc\@ifnch}

\def\@ifnextchar#1#2#3{%
  \let\@tempMACe #1%
  \def\@tempMACa{#2}%
  \def\@tempMACb{#3}%
  \futurelet \@tempMACc\@ifnch%
}

\def\@ifnch{%
\ifx \@tempMACc \@sptoken%
  \let\@tempMACd\@xifnch%
\else%
  \ifx \@tempMACc \@tempMACe%
    \let\@tempMACd\@tempMACa%
  \else%
    \let\@tempMACd\@tempMACb%
  \fi%
\fi%
\@tempMACd%
}

\def\@ifstar#1#2{\@ifnextchar *{\def\@tempMACa*{#1}\@tempMACa}{#2}}

\def\section{\@ifstar{\@ssection}{\@section}}

\def\@section#1{
 \EveryMac
 \Two
 \LastMac=\Hae
 \tenpoint\bls{\ElevenPT}
 \bf
 \Raggedright
 \ifAutoNumber
  \advance\Sec by 1
  \noindent\number\Sec\hskip 1pc \uppercase{#1}
  \SecSec=0
 \else
  \noindent \uppercase{#1}
 \fi
 \nobreak
}

\def\@ssection#1{
 \EveryMac
 \ifnum\LastMac=\Hae \Half
  \else \OneHalf
 \fi
 \LastMac=\Hae
 \tenpoint\bls{\ElevenPT}
 \bf
 \Raggedright
 \noindent\uppercase{#1}
}

\def\subsection#1{
 \EveryMac
 \ifnum\LastMac=\Hae \Half
  \else \OneHalf
 \fi
 \LastMac=\Hbe
 \tenpoint\bls{\ElevenPT}
 \bf
 \Raggedright
 \ifAutoNumber
  \advance\SecSec by 1
  \noindent\number\Sec.\number\SecSec
  \hskip 1pc #1
  \SecSecSec=0
 \else
  \noindent #1
 \fi
 \nobreak
}

\def\subsubsection#1{
 \EveryMac
 \ifnum\LastMac=\Hbe \Half
  \else \OneHalf
 \fi
 \LastMac=\Hce
 \ninepoint\bls{\ElevenPT}
 \it
 \Raggedright
 \ifAutoNumber
  \advance\SecSecSec by 1
  \noindent\number\Sec.\number\SecSec.\number\SecSecSec
  \hskip 1pc #1
 \else
  \noindent #1
 \fi
 \nobreak
}

\def\paragraph#1{
 \EveryMac
 \One
 \LastMac=\Hde
 \ninepoint\bls{\ElevenPT}
 \noindent \it #1
 \rm
}


\def\tx{
 \EveryMac
 \ifnum\LastMac=\Lie \Half\fi
 \ifnum\LastMac=\Hae \nobreak\Half\fi
 \ifnum\LastMac=\Hbe \nobreak\Half\fi
 \ifnum\LastMac=\Hce \nobreak\Half\fi
 \ifnum\LastMac=\Lie \else \noindent\fi
 \LastMac=\Txe
 \ninepoint\bls{\ElevenPT}
 \rm
}


\def\item{
 \par
 \EveryMac
 \ifnum\LastMac=\Lie
  \else \Half
 \fi
 \LastMac=\Lie
 \ninepoint\bls{\ElevenPT}
 \rm
}


\def\bibitem{
 \par
 \EveryMac
 \ifnum\LastMac=\Bbe
  \else \Half
 \fi
 \LastMac=\Bbe
 \Hang{1.5em}{1}
 \eightpoint\bls{\TenPT}
 \Raggedright
 \noindent \rm
}


\newtoks\CatchLine

\def\@journal{Mon.\ Not.\ R.\ Astron.\ Soc.\ }  
\def\@pubyear{1994}        
\def\@pagerange{000--000}  
\def\@volume{000}          
\def\@microfiche{}         %

\def\pubyear#1{\gdef\@pubyear{#1}\@makecatchline}
\def\pagerange#1{\gdef\@pagerange{#1}\@makecatchline}
\def\volume#1{\gdef\@volume{#1}\@makecatchline}
\def\microfiche#1{\gdef\@microfiche{and Microfiche\ #1}\@makecatchline}

\def\@makecatchline{%
  \global\CatchLine{%
    {\rm \@journal {\bf \@volume},\ \@pagerange\ (\@pubyear)\ \@microfiche}}%
}

\@makecatchline 

\newtoks\LeftHeader
\def\shortauthor#1{
 \global\LeftHeader{#1}
}

\newtoks\RightHeader
\def\shorttitle#1{
 \global\RightHeader{#1}
}

\def\PageHead{
 \EveryMac
 \ifnum\HeaderNumber=1 \Pagehead
  \else \Catchline
 \fi
}

\def\Catchline{%
 \vbox to 0pt{\vskip-22.5pt
  \hbox to \PageWidth{\vbox to8.5pt{}\noindent
  \eightpoint\the\CatchLine\hfill}\vss}
 \nointerlineskip
}

\def\Pagehead{%
 \ifodd\pageno
   \vbox to 0pt{\vskip-22.5pt
   \hbox to \PageWidth{\vbox to8.5pt{}\elevenpoint\it\noindent
    \hfill\the\RightHeader\hskip1.5em\rm\folio}\vss}
 \else
   \vbox to 0pt{\vskip-22.5pt
   \hbox to \PageWidth{\vbox to8.5pt{}\elevenpoint\rm\noindent
   \folio\hskip1.5em\it\the\LeftHeader\hfill}\vss}
 \fi
 \nointerlineskip
}

\def\PageFoot{} 

\def\authorcomment#1{%
  \gdef\PageFoot{%
    \nointerlineskip%
    \vbox to 22pt{\vfil%
      \hbox to \PageWidth{\elevenpoint\rm\noindent \hfil #1 \hfil}}%
  }%
}

\everydisplay{\displaysetup}

\newif\ifeqno
\newif\ifleqno

\def\displaysetup#1$${%
 \displaytest#1\eqno\eqno\displaytest
}

\def\displaytest#1\eqno#2\eqno#3\displaytest{%
 \if!#3!\ldisplaytest#1\leqno\leqno\ldisplaytest
 \else\eqnotrue\leqnofalse\def\eqn{#2}\def\eq{#1}\fi
 \generaldisplay$$}

\def\ldisplaytest#1\leqno#2\leqno#3\ldisplaytest{%
 \def\eq{#1}%
 \if!#3!\eqnofalse\else\eqnotrue\leqnotrue
  \def\eqn{#2}\fi}

\def\generaldisplay{%
\ifeqno \ifleqno 
   \hbox to \hsize{\noindent
     $\displaystyle\eq$\hfil$\displaystyle\eqn$}
  \else
    \hbox to \hsize{\noindent
     $\displaystyle\eq$\hfil$\displaystyle\eqn$}
  \fi
 \else
 \hbox to \hsize{\vbox{\noindent
  $\displaystyle\eq$\hfil}}
 \fi
}

\def\@notice{%
  \par\Two%
  \bls{12pt}%
  \noindent\tenrm This paper has been produced using the Blackwell
                  Scientific Publications \TeX\ macros.%
}

\outer\def\bye{\@notice\par\vfill\supereject\end}

\everyjob{%
  \Warn{Monthly notices of the RAS journal style (\@typeface)\space
        v\@version,\space \@verdate.}\Warn{}%
}




\newif\if@debug \@debugfalse  

\def\Print#1{\if@debug\immediate\write16{#1}\else \fi}
\def\Warn#1{\immediate\write16{#1}}
\def\wlog#1{}

\newcount\Iteration 

\newif\ifFigureBoxes  
\FigureBoxestrue

\def\Single{0} \def\Double{1}                 
\def\Figure{0} \def\Table{1}                  

\def\InStack{0}  
\def\InZoneA{1}
\def\InZoneB{2}
\def\InZoneC{3}

\newcount\TEMPCOUNT 
\newdimen\TEMPDIMEN 
\newbox\TEMPBOX     
\newbox\VOIDBOX     

\newcount\LengthOfStack 
\newcount\MaxItems      
\newcount\StackPointer
\newcount\Point         
\newcount\NextFigure    
\newcount\NextTable     
\newcount\NextItem      

\newcount\StatusStack   
\newcount\NumStack      
\newcount\TypeStack     
\newcount\SpanStack     
\newcount\BoxStack      

\newcount\ItemSTATUS    
\newcount\ItemNUMBER    
\newcount\ItemTYPE      
\newcount\ItemSPAN      
\newbox\ItemBOX         
\newdimen\ItemSIZE      

\newdimen\PageHeight    
\newdimen\TextLeading   
\newdimen\Feathering    
\newcount\LinesPerPage  
\newdimen\ColumnWidth   
\newdimen\ColumnGap     
\newdimen\PageWidth     
\newdimen\BodgeHeight   
\newcount\Leading       

\newdimen\ZoneBSize  
\newdimen\TextSize   
\newbox\ZoneABOX     
\newbox\ZoneBBOX     
\newbox\ZoneCBOX     

\newif\ifFirstSingleItem
\newif\ifFirstZoneA
\newif\ifMakePageInComplete
\newif\ifMoreFigures \MoreFiguresfalse 
\newif\ifMoreTables  \MoreTablesfalse  

\newif\ifFigInZoneB 
\newif\ifFigInZoneC 
\newif\ifTabInZoneB 
\newif\ifTabInZoneC

\newif\ifZoneAFullPage

\newbox\MidBOX    
\newbox\LeftBOX
\newbox\RightBOX
\newbox\PageBOX   

\newif\ifLeftCOL  
\LeftCOLtrue

\newdimen\ZoneBAdjust

\newcount\ItemFits
\def\Yes{1}
\def\No{2}




\MaxItems=15
\NextFigure=0        
\NextTable=1

\BodgeHeight=6pt
\TextLeading=11pt    
\Leading=11
\Feathering=0pt      
\LinesPerPage=61     
\topskip=\TextLeading
\ColumnWidth=20pc    
\ColumnGap=2pc       

\def\ItemSep{\vskip \TextLeading plus \TextLeading minus 4pt}

\FigureBoxesfalse 

\parskip=0pt
\parindent=18pt
\widowpenalty=0
\clubpenalty=10000
\tolerance=1500
\hbadness=1500
\abovedisplayskip=6pt plus 2pt minus 2pt
\belowdisplayskip=6pt plus 2pt minus 2pt
\abovedisplayshortskip=6pt plus 2pt minus 2pt
\belowdisplayshortskip=6pt plus 2pt minus 2pt

\PageHeight=\TextLeading 
\multiply\PageHeight by \LinesPerPage
\advance\PageHeight by \topskip

\PageWidth=2\ColumnWidth
\advance\PageWidth by \ColumnGap




\newcount\DUMMY \StatusStack=\allocationnumber
\newcount\DUMMY \newcount\DUMMY \newcount\DUMMY 
\newcount\DUMMY \newcount\DUMMY \newcount\DUMMY 
\newcount\DUMMY \newcount\DUMMY \newcount\DUMMY
\newcount\DUMMY \newcount\DUMMY \newcount\DUMMY 
\newcount\DUMMY \newcount\DUMMY \newcount\DUMMY

\newcount\DUMMY \NumStack=\allocationnumber
\newcount\DUMMY \newcount\DUMMY \newcount\DUMMY 
\newcount\DUMMY \newcount\DUMMY \newcount\DUMMY 
\newcount\DUMMY \newcount\DUMMY \newcount\DUMMY 
\newcount\DUMMY \newcount\DUMMY \newcount\DUMMY 
\newcount\DUMMY \newcount\DUMMY \newcount\DUMMY

\newcount\DUMMY \TypeStack=\allocationnumber
\newcount\DUMMY \newcount\DUMMY \newcount\DUMMY 
\newcount\DUMMY \newcount\DUMMY \newcount\DUMMY 
\newcount\DUMMY \newcount\DUMMY \newcount\DUMMY 
\newcount\DUMMY \newcount\DUMMY \newcount\DUMMY 
\newcount\DUMMY \newcount\DUMMY \newcount\DUMMY

\newcount\DUMMY \SpanStack=\allocationnumber
\newcount\DUMMY \newcount\DUMMY \newcount\DUMMY 
\newcount\DUMMY \newcount\DUMMY \newcount\DUMMY 
\newcount\DUMMY \newcount\DUMMY \newcount\DUMMY 
\newcount\DUMMY \newcount\DUMMY \newcount\DUMMY 
\newcount\DUMMY \newcount\DUMMY \newcount\DUMMY

\newbox\DUMMY   \BoxStack=\allocationnumber
\newbox\DUMMY   \newbox\DUMMY \newbox\DUMMY 
\newbox\DUMMY   \newbox\DUMMY \newbox\DUMMY 
\newbox\DUMMY   \newbox\DUMMY \newbox\DUMMY 
\newbox\DUMMY   \newbox\DUMMY \newbox\DUMMY 
\newbox\DUMMY   \newbox\DUMMY \newbox\DUMMY

\def\wlog{\immediate\write-1}


\def\GetItemAll#1{%
 \GetItemSTATUS{#1}
 \GetItemNUMBER{#1}
 \GetItemTYPE{#1}
 \GetItemSPAN{#1}
 \GetItemBOX{#1}
}

\def\GetItemSTATUS#1{%
 \Point=\StatusStack
 \advance\Point by #1
 \global\ItemSTATUS=\count\Point
}

\def\GetItemNUMBER#1{%
 \Point=\NumStack
 \advance\Point by #1
 \global\ItemNUMBER=\count\Point
}

\def\GetItemTYPE#1{%
 \Point=\TypeStack
 \advance\Point by #1
 \global\ItemTYPE=\count\Point
}

\def\GetItemSPAN#1{%
 \Point\SpanStack
 \advance\Point by #1
 \global\ItemSPAN=\count\Point
}

\def\GetItemBOX#1{%
 \Point=\BoxStack
 \advance\Point by #1
 \global\setbox\ItemBOX=\vbox{\copy\Point}
 \global\ItemSIZE=\ht\ItemBOX
 \global\advance\ItemSIZE by \dp\ItemBOX
 \TEMPCOUNT=\ItemSIZE
 \divide\TEMPCOUNT by \Leading
 \divide\TEMPCOUNT by 65536
 \advance\TEMPCOUNT by 1
 \ItemSIZE=\TEMPCOUNT pt
 \global\multiply\ItemSIZE by \Leading
}


\def\JoinStack{%
 \ifnum\LengthOfStack=\MaxItems 
  \Warn{WARNING: Stack is full...some items will be lost!}
 \else
  \Point=\StatusStack
  \advance\Point by \LengthOfStack
  \global\count\Point=\ItemSTATUS
  \Point=\NumStack
  \advance\Point by \LengthOfStack
  \global\count\Point=\ItemNUMBER
  \Point=\TypeStack
  \advance\Point by \LengthOfStack
  \global\count\Point=\ItemTYPE
  \Point\SpanStack
  \advance\Point by \LengthOfStack
  \global\count\Point=\ItemSPAN
  \Point=\BoxStack
  \advance\Point by \LengthOfStack
  \global\setbox\Point=\vbox{\copy\ItemBOX}
  \global\advance\LengthOfStack by 1
  \ifnum\ItemTYPE=\Figure 
   \global\MoreFigurestrue
  \else
   \global\MoreTablestrue
  \fi
 \fi
}


\def\LeaveStack#1{%
 {\Iteration=#1
 \loop
 \ifnum\Iteration<\LengthOfStack
  \advance\Iteration by 1
  \GetItemSTATUS{\Iteration}
   \advance\Point by -1
   \global\count\Point=\ItemSTATUS
  \GetItemNUMBER{\Iteration}
   \advance\Point by -1
   \global\count\Point=\ItemNUMBER
  \GetItemTYPE{\Iteration}
   \advance\Point by -1
   \global\count\Point=\ItemTYPE
  \GetItemSPAN{\Iteration}
   \advance\Point by -1
   \global\count\Point=\ItemSPAN
  \GetItemBOX{\Iteration}
   \advance\Point by -1
   \global\setbox\Point=\vbox{\copy\ItemBOX}
 \repeat}
 \global\advance\LengthOfStack by -1
}


\newif\ifStackNotClean

\def\CleanStack{%
 \StackNotCleantrue
 {\Iteration=0
  \loop
   \ifStackNotClean
    \GetItemSTATUS{\Iteration}
    \ifnum\ItemSTATUS=\InStack
     \advance\Iteration by 1
     \else
      \LeaveStack{\Iteration}
    \fi
   \ifnum\LengthOfStack<\Iteration
    \StackNotCleanfalse
   \fi
 \repeat}
}


\def\FindItem#1#2{%
 \global\StackPointer=-1 
 {\Iteration=0
  \loop
  \ifnum\Iteration<\LengthOfStack
   \GetItemSTATUS{\Iteration}
   \ifnum\ItemSTATUS=\InStack
    \GetItemTYPE{\Iteration}
    \ifnum\ItemTYPE=#1
     \GetItemNUMBER{\Iteration}
     \ifnum\ItemNUMBER=#2
      \global\StackPointer=\Iteration
      \Iteration=\LengthOfStack 
     \fi
    \fi
   \fi
  \advance\Iteration by 1
 \repeat}
}


\def\FindNext{%
 \global\StackPointer=-1 
 {\Iteration=0
  \loop
  \ifnum\Iteration<\LengthOfStack
   \GetItemSTATUS{\Iteration}
   \ifnum\ItemSTATUS=\InStack
    \GetItemTYPE{\Iteration}
   \ifnum\ItemTYPE=\Figure
    \ifMoreFigures
      \global\NextItem=\Figure
      \global\StackPointer=\Iteration
      \Iteration=\LengthOfStack 
    \fi
   \fi
   \ifnum\ItemTYPE=\Table
    \ifMoreTables
      \global\NextItem=\Table
      \global\StackPointer=\Iteration
      \Iteration=\LengthOfStack 
    \fi
   \fi
  \fi
  \advance\Iteration by 1
 \repeat}
}


\def\ChangeStatus#1#2{%
 \Point=\StatusStack
 \advance\Point by #1
 \global\count\Point=#2
}



\def\Zone{\InZoneA}

\ZoneBAdjust=0pt

\def\MakePage{
 \global\ZoneBSize=\PageHeight
 \global\TextSize=\ZoneBSize
 \global\ZoneAFullPagefalse
 \global\topskip=\TextLeading
 \MakePageInCompletetrue
 \MoreFigurestrue
 \MoreTablestrue
 \FigInZoneBfalse
 \FigInZoneCfalse
 \TabInZoneBfalse
 \TabInZoneCfalse
 \global\FirstSingleItemtrue
 \global\FirstZoneAtrue
 \global\setbox\ZoneABOX=\box\VOIDBOX
 \global\setbox\ZoneBBOX=\box\VOIDBOX
 \global\setbox\ZoneCBOX=\box\VOIDBOX
 \loop
  \ifMakePageInComplete
 \FindNext
 \ifnum\StackPointer=-1
  \NextItem=-1
  \MoreFiguresfalse
  \MoreTablesfalse
 \fi
 \ifnum\NextItem=\Figure
   \FindItem{\Figure}{\NextFigure}
   \ifnum\StackPointer=-1 \global\MoreFiguresfalse
   \else
    \GetItemSPAN{\StackPointer}
    \ifnum\ItemSPAN=\Single \def\Zone{\InZoneB}\relax
     \ifFigInZoneC \global\MoreFiguresfalse\fi
    \else
     \def\Zone{\InZoneA}
     \ifFigInZoneB \def\Zone{\InZoneC}\fi
    \fi
   \fi
   \ifMoreFigures\Print{}\FigureItems\fi
 \fi
\ifnum\NextItem=\Table
   \FindItem{\Table}{\NextTable}
   \ifnum\StackPointer=-1 \global\MoreTablesfalse
   \else
    \GetItemSPAN{\StackPointer}
    \ifnum\ItemSPAN=\Single\relax
     \ifTabInZoneC \global\MoreTablesfalse\fi
    \else
     \def\Zone{\InZoneA}
     \ifTabInZoneB \def\Zone{\InZoneC}\fi
    \fi
   \fi
   \ifMoreTables\Print{}\TableItems\fi
 \fi
   \MakePageInCompletefalse 
   \ifMoreFigures\MakePageInCompletetrue\fi
   \ifMoreTables\MakePageInCompletetrue\fi
 \repeat
 \ifZoneAFullPage
  \global\TextSize=0pt
  \global\ZoneBSize=0pt
  \global\vsize=0pt\relax
  \global\topskip=0pt\relax
  \vbox to 0pt{\vss}
  \eject
 \else
 \global\advance\ZoneBSize by -\ZoneBAdjust
 \global\vsize=\ZoneBSize
 \global\hsize=\ColumnWidth
 \global\ZoneBAdjust=0pt
 \ifdim\TextSize<23pt
 \Warn{}
 \Warn{* Making column fall short: TextSize=\the\TextSize *}
 \vskip-\lastskip\eject\fi
 \fi
}

\def\MakeRightCol{
 \global\TextSize=\ZoneBSize
 \MakePageInCompletetrue
 \MoreFigurestrue
 \MoreTablestrue
 \global\FirstSingleItemtrue
 \global\setbox\ZoneBBOX=\box\VOIDBOX
 \def\Zone{\InZoneB}
 \loop
  \ifMakePageInComplete
 \FindNext
 \ifnum\StackPointer=-1
  \NextItem=-1
  \MoreFiguresfalse
  \MoreTablesfalse
 \fi
 \ifnum\NextItem=\Figure
   \FindItem{\Figure}{\NextFigure}
   \ifnum\StackPointer=-1 \MoreFiguresfalse
   \else
    \GetItemSPAN{\StackPointer}
    \ifnum\ItemSPAN=\Double\relax
     \MoreFiguresfalse\fi
   \fi
   \ifMoreFigures\Print{}\FigureItems\fi
 \fi
 \ifnum\NextItem=\Table
   \FindItem{\Table}{\NextTable}
   \ifnum\StackPointer=-1 \MoreTablesfalse
   \else
    \GetItemSPAN{\StackPointer}
    \ifnum\ItemSPAN=\Double\relax
     \MoreTablesfalse\fi
   \fi
   \ifMoreTables\Print{}\TableItems\fi
 \fi
   \MakePageInCompletefalse 
   \ifMoreFigures\MakePageInCompletetrue\fi
   \ifMoreTables\MakePageInCompletetrue\fi
 \repeat
 \ifZoneAFullPage
  \global\TextSize=0pt
  \global\ZoneBSize=0pt
  \global\vsize=0pt\relax
  \global\topskip=0pt\relax
  \vbox to 0pt{\vss}
  \eject
 \else
 \global\vsize=\ZoneBSize
 \global\hsize=\ColumnWidth
 \ifdim\TextSize<23pt
 \Warn{}
 \Warn{* Making column fall short: TextSize=\the\TextSize *}
 \vskip-\lastskip\eject\fi
\fi
}

\def\FigureItems{
 \Print{Considering...}
 \ShowItem{\StackPointer}
 \GetItemBOX{\StackPointer} 
 \GetItemSPAN{\StackPointer}
  \CheckFitInZone 
  \ifnum\ItemFits=\Yes
   \ifnum\ItemSPAN=\Single
     \ChangeStatus{\StackPointer}{\InZoneB} 
     \global\FigInZoneBtrue
     \ifFirstSingleItem
      \hbox{}\vskip-\BodgeHeight
     \global\advance\ItemSIZE by \TextLeading
     \fi
     \unvbox\ItemBOX\ItemSep
     \global\FirstSingleItemfalse
     \global\advance\TextSize by -\ItemSIZE
     \global\advance\TextSize by -\TextLeading
   \else
    \ifFirstZoneA
     \global\advance\ItemSIZE by \TextLeading
     \global\FirstZoneAfalse\fi
    \global\advance\TextSize by -\ItemSIZE
    \global\advance\TextSize by -\TextLeading
    \global\advance\ZoneBSize by -\ItemSIZE
    \global\advance\ZoneBSize by -\TextLeading
    \ifFigInZoneB\relax
     \else
     \ifdim\TextSize<3\TextLeading
     \global\ZoneAFullPagetrue
     \fi
    \fi
    \ChangeStatus{\StackPointer}{\Zone}
    \ifnum\Zone=\InZoneC \global\FigInZoneCtrue\fi
  \fi
   \Print{TextSize=\the\TextSize}
   \Print{ZoneBSize=\the\ZoneBSize}
  \global\advance\NextFigure by 1
   \Print{This figure has been placed.}
  \else
   \Print{No space available for this figure...holding over.}
   \Print{}
   \global\MoreFiguresfalse
  \fi
}

\def\TableItems{
 \Print{Considering...}
 \ShowItem{\StackPointer}
 \GetItemBOX{\StackPointer} 
 \GetItemSPAN{\StackPointer}
  \CheckFitInZone 
  \ifnum\ItemFits=\Yes
   \ifnum\ItemSPAN=\Single
    \ChangeStatus{\StackPointer}{\InZoneB}
     \global\TabInZoneBtrue
     \ifFirstSingleItem
      \hbox{}\vskip-\BodgeHeight
     \global\advance\ItemSIZE by \TextLeading
     \fi
     \unvbox\ItemBOX\ItemSep
     \global\FirstSingleItemfalse
     \global\advance\TextSize by -\ItemSIZE
     \global\advance\TextSize by -\TextLeading
   \else
    \ifFirstZoneA
    \global\advance\ItemSIZE by \TextLeading
    \global\FirstZoneAfalse\fi
    \global\advance\TextSize by -\ItemSIZE
    \global\advance\TextSize by -\TextLeading
    \global\advance\ZoneBSize by -\ItemSIZE
    \global\advance\ZoneBSize by -\TextLeading
    \ifFigInZoneB\relax
     \else
     \ifdim\TextSize<3\TextLeading
     \global\ZoneAFullPagetrue
     \fi
    \fi
    \ChangeStatus{\StackPointer}{\Zone}
    \ifnum\Zone=\InZoneC \global\TabInZoneCtrue\fi
   \fi
  \global\advance\NextTable by 1
   \Print{This table has been placed.}
  \else
  \Print{No space available for this table...holding over.}
   \Print{}
   \global\MoreTablesfalse
  \fi
}


\def\CheckFitInZone{%
{\advance\TextSize by -\ItemSIZE
 \advance\TextSize by -\TextLeading
 \ifFirstSingleItem
  \advance\TextSize by \TextLeading
 \fi
 \ifnum\Zone=\InZoneA\relax
  \else \advance\TextSize by -\ZoneBAdjust
 \fi
 \ifdim\TextSize<3\TextLeading \global\ItemFits=\No
 \else \global\ItemFits=\Yes\fi}
}

\def\BF#1#2{
 \ItemSTATUS=\InStack
 \ItemNUMBER=#1
 \ItemTYPE=\Figure
 \if#2S \ItemSPAN=\Single
  \else \ItemSPAN=\Double
 \fi
 \setbox\ItemBOX=\vbox{}
}

\def\BT#1#2{
 \ItemSTATUS=\InStack
 \ItemNUMBER=#1
 \ItemTYPE=\Table
 \if#2S \ItemSPAN=\Single
  \else \ItemSPAN=\Double
 \fi
 \setbox\ItemBOX=\vbox{}
}

\def\BeginOpening{%
 \hsize=\PageWidth
 \global\setbox\ItemBOX=\vbox\bgroup
}

\let\begintopmatter=\BeginOpening  

\def\EndOpening{%
 \egroup
 \ItemNUMBER=0
 \ItemTYPE=\Figure
 \ItemSPAN=\Double
 \ItemSTATUS=\InStack
 \JoinStack
}


\newbox\tmpbox

\def\FC#1#2#3#4{%
  \ItemSTATUS=\InStack
  \ItemNUMBER=#1
  \ItemTYPE=\Figure
  \if#2S
    \ItemSPAN=\Single \TEMPDIMEN=\ColumnWidth
  \else
    \ItemSPAN=\Double \TEMPDIMEN=\PageWidth
  \fi
  {\hsize=\TEMPDIMEN
   \global\setbox\ItemBOX=\vbox{%
     \ifFigureBoxes
       \B{\TEMPDIMEN}{#3}
     \else
       \vbox to #3{\vfil}%
     \fi%
     \eightpoint\rm\bls{\rTenPT}%
     \vskip 5.5pt plus 6pt%
     \setbox\tmpbox=\vbox{#4\par}%
     \ifdim\ht\tmpbox>10pt 
       \noindent #4\par%
     \else
       \hbox to \hsize{\hfil #4\hfil}%
     \fi%
   }%
  }%
  \JoinStack%
  \Print{Processing source for figure {\the\ItemNUMBER}}%
}


\def\figps#1#2#3#4{%
  \ItemSTATUS=\InStack
  \ItemNUMBER=#1
  \ItemTYPE=\Figure
  \if#2S
    \ItemSPAN=\Single \TEMPDIMEN=\ColumnWidth
  \else
    \ItemSPAN=\Double \TEMPDIMEN=\PageWidth
  \fi
  {\hsize=\TEMPDIMEN
   \global\setbox\ItemBOX=\vbox{#3
     \eightpoint\rm\bls{\rTenPT}%
     \vskip 5.5pt plus 6pt%
     \setbox\tmpbox=\vbox{#4\par}%
     \ifdim\ht\tmpbox>10pt 
       \noindent #4\par%
     \else
       \hbox to \hsize{\hfil #4\hfil}%
     \fi%
   }%
  }%
  \JoinStack%
  \Print{Processing source for figure {\the\ItemNUMBER}}%
}

\def\TH#1#2#3#4{%
 \ItemSTATUS=\InStack
 \ItemNUMBER=#1
 \ItemTYPE=\Table
 \if#2S \ItemSPAN=\Single \TEMPDIMEN=\ColumnWidth
  \else \ItemSPAN=\Double \TEMPDIMEN=\PageWidth
 \fi
 {\hsize=\TEMPDIMEN
 \eightpoint\bls{\rTenPT}\rm
 \global\setbox\ItemBOX=\vbox{\noindent#3\vskip 5.5pt plus5.5pt\noindent#4}}
 \JoinStack
 \Print{Processing source for table {\the\ItemNUMBER}}
}

\let\table=\TH  

\def\UnloadZoneA{%
\FirstZoneAtrue
 \Iteration=0
  \loop
   \ifnum\Iteration<\LengthOfStack
    \GetItemSTATUS{\Iteration}
    \ifnum\ItemSTATUS=\InZoneA
     \GetItemBOX{\Iteration}
     \ifFirstZoneA \vbox to \BodgeHeight{\vfil}%
     \FirstZoneAfalse\fi
     \unvbox\ItemBOX\ItemSep
     \LeaveStack{\Iteration}
     \else
     \advance\Iteration by 1
   \fi
 \repeat
}

\def\UnloadZoneC{%
\Iteration=0
  \loop
   \ifnum\Iteration<\LengthOfStack
    \GetItemSTATUS{\Iteration}
    \ifnum\ItemSTATUS=\InZoneC
     \GetItemBOX{\Iteration}
     \ItemSep\unvbox\ItemBOX
     \LeaveStack{\Iteration}
     \else
     \advance\Iteration by 1
   \fi
 \repeat
}


\def\ShowItem#1{
  {\GetItemAll{#1}
  \Print{\the#1:
  {TYPE=\ifnum\ItemTYPE=\Figure Figure\else Table\fi}
  {NUMBER=\the\ItemNUMBER}
  {SPAN=\ifnum\ItemSPAN=\Single Single\else Double\fi}
  {SIZE=\the\ItemSIZE}}}
}

\def\ShowStack{%
 \Print{}
 \Print{LengthOfStack = \the\LengthOfStack}
 \ifnum\LengthOfStack=0 \Print{Stack is empty}\fi
 \Iteration=0
 \loop
 \ifnum\Iteration<\LengthOfStack
  \ShowItem{\Iteration}
  \advance\Iteration by 1
 \repeat
}

\def\B#1#2{%
\hbox{\vrule\kern-0.4pt\vbox to #2{%
\hrule width #1\vfill\hrule}\kern-0.4pt\vrule}
}

\def\Ref#1{\begingroup\global\setbox\TEMPBOX=\vbox{\hsize=2in\noindent#1}\endgroup
\ht1=0pt\dp1=0pt\wd1=0pt\vadjust{\vtop to 0pt{\advance
\hsize0.5pc\kern-10pt\moveright\hsize\box\TEMPBOX\vss}}}

\def\MarkRef#1{\leavevmode\thinspace\hbox{\vrule\vtop
{\vbox{\hrule\kern1pt\hbox{\vphantom{\rm/}\thinspace{\rm#1}%
\thinspace}}\kern1pt\hrule}\vrule}\thinspace}%


\output{%
 \ifLeftCOL
  \global\setbox\LeftBOX=\vbox to \ZoneBSize{\box255\unvbox\ZoneBBOX}
  \global\LeftCOLfalse
  \MakeRightCol
 \else
  \setbox\RightBOX=\vbox to \ZoneBSize{\box255\unvbox\ZoneBBOX}
  \setbox\MidBOX=\hbox{\box\LeftBOX\hskip\ColumnGap\box\RightBOX}
  \setbox\PageBOX=\vbox to \PageHeight{%
  \UnloadZoneA\box\MidBOX\UnloadZoneC}
  \shipout\vbox{\PageHead\box\PageBOX\PageFoot}
  \global\advance\pageno by 1
  \global\HeaderNumber=\DefaultHeader
  \global\LeftCOLtrue
  \CleanStack
  \MakePage
 \fi
}


\catcode `\@=12 




\catcode`@=11
\def\note#1#2{%
  \let\@sf=\empty \ifhmode\edef\@sf{\spacefactor\the\spacefactor}\/\fi
  $\m@th{}^{#1}$%
  \insert\footins\bgroup
    \eightpoint\bls{\rTenPT}\rm
    \textindent{$\m@th{}^{#1}$}%
    \interlinepenalty\interfootnotelinepenalty
    \splittopskip\ht\strutbox 
    \splitmaxdepth\dp\strutbox \floatingpenalty\@MM
    \leftskip\z@skip \rightskip\z@skip \spaceskip\z@skip \xspaceskip\z@skip
    \footstrut #2\strut\par
  \egroup
  \@sf\relax}
 

\output{%
 \ifLeftCOL
  \global\setbox\LeftBOX=\vbox to \ZoneBSize{\box255\unvbox\ZoneBBOX
    \ifvoid\footins\else
    \vskip\skip\footins\unvbox\footins\fi}
  \global\LeftCOLfalse
  \MakeRightCol
 \else
  \setbox\RightBOX=\vbox to \ZoneBSize{\box255\unvbox\ZoneBBOX
    \ifvoid\footins\else
    \vskip\skip\footins\unvbox\footins\fi}
  \setbox\MidBOX=\hbox{\box\LeftBOX\hskip\ColumnGap\box\RightBOX}
  \setbox\PageBOX=\vbox to \PageHeight{%
  \UnloadZoneA\box\MidBOX\UnloadZoneC}
  \shipout\vbox{\PageHead\box\PageBOX\PageFoot}
  \global\advance\pageno by 1
  \global\HeaderNumber=\DefaultHeader
  \global\LeftCOLtrue
  \CleanStack
  \MakePage
 \fi
}

\input psfig

\newcount\eqnumber
\eqnumber=1
\def\step#1{\global\advance#1 by 1}

\def\neweq{{\rm\the\eqnumber}\step{\eqnumber}}
\def\preveq#1){\advance\eqnumber by -#1 \the\eqnumber) \advance\eqnumber by #1}

\pageoffset{-2.5pc}{0pc}

\Autonumber  
\pagerange{000--000}
\pubyear{1998}
\volume{}

\begintopmatter  
\title{Overmerging and M/L ratios in phenomenological galaxy formation models} 
\author{Eelco van Kampen$^{1,2}$, Raul Jimenez$^1$ and John A. Peacock$^1$}
\affiliation{$^1$ Institute for Astronomy, University of Edinburgh,
Royal Observatory, Blackford Hill, Edinburgh EH9 3HJ}
\affiliation{$^2$ Theoretical Astrophysics Center, Juliane Maries Vej 30,
DK-2100 Copenhagen {\O}, Denmark}

\shortauthor{van Kampen et al.}
\shorttitle{phenomenological galaxy formation}
\acceptedline{Accepted ... Received ...; in original form ...}
\abstract We show that the discrepancy between the Tully-Fisher relation
and the luminosity function predicted by most phenomenological
galaxy formation models is mainly due to overmerging of galaxy haloes.
We have circumvented this overmerging problem, which is inherent in both
the Press-Schechter formalism and dissipationless N-body simulations,
by including a specific galaxy halo formation recipe into an
otherwise standard N-body code. This numerical technique provides
the merger trees which, together with simplified gas dynamics and
star formation physics, constitute our implementation of a
phenomenological galaxy formation model. Resolving the overmerging
problem provides us with the means to match both the I-band
Tully-Fisher relation and the B and K band luminosity functions
within an $\Omega=1$ sCDM structure formation scenario.
It also allows us to include models for chemical evolution and
starbursts, which improves the match to observational data {\it and}
renders the modelling more realistic. We show that the inclusion of
chemical evolution into the modelling requires a significant
fraction of stars to be formed in short bursts triggered by merging
events.

\keywords cosmology: theory - dark matter - large-scale structure of Universe
- galaxy formation
\maketitle   

\section{Introduction}

\tx There has been significant recent progress in the study of galaxy
formation within a cosmological context, mainly due to a phenomenological
or `semi-analytical' approach to this problem. The idea is to start with
a structure formation model that describes where and when galactic
dark haloes form. A simple description of gas dynamics and star formation
provides a means to calculate the amount of stars forming in these haloes.
Stellar population synthesis models then provide the spectral evolution,
i.e.\ luminosities and colours, of these galaxies.

Many physical processes are modelled as simple functions of the circular
velocity of the galaxy halo. Therefore, the Tully-Fisher relation is
the most obvious observational relation to try and predict, as it
relates the total luminosity of a galaxy to its halo circular
velocity. However, most phenomenological galaxy formation models do
not simultaneously fit the I-band Tully-Fisher relation and the
B or K band luminosity function. When one sets the model parameters
such that the Tully-Fisher relation has the right normalization, the
luminosity functions generally overshoot (e.g.\ Kauffman, White \&
Guiderdoni 1993; Kaufmann, Colberg, Diaferio \& White 1998), certainly
for the $\Omega=1$, $H_0=50$ km s$^{-1}$ Mpc$^{-1}$
standard CDM cosmology (in the form given by Davis et al.\ 1985) that
we consider in this paper. Alternatively, when making sure that the
luminosity functions matches by changing some of the model parameters,
the Tully-Fisher relation ends up significantly shifted with respect to
the observed relation (e.g.\ Cole et al.\ 1994; Heyl et al.\ 1995).

In order to keep the modelling as analytical as possible, an extension
to the Press \& Schechter (1974) prescription for the evolution of galaxy
haloes (e.g.\ Bond et al. 1991; Bower 1991; Lacey \& Cole 1993; Kauffman
\& White 1993) has been a popular ingredient for implementations of a
phenomenological theory of galaxy formation. The extended Press-Schechter
(EPS) formalism keeps the galaxy formation model as analytical as possible,
and allows fast realizations of halo populations. 
However, this prescription has the disadvantage that the large-scale
phase-space distribution of haloes is unknown, as is any information on
their internal structure and kinematics.
For many applications this information will be of great use.
Furthermore, the properties and formation histories of {\it individual}\
haloes do not match the corresponding properties and histories
as predicted by the EPS formalism (e.g.\ Lacey \& Cole 1993; White 1996).

More importantly, the EPS formalism is designed to identify collapsed
systems, irrespective of whether these contain surviving subsystems.
This `overmerging' of subhaloes into larger embedding haloes produces
a top-heavy galaxy halo mass function, and is therefore
relevant to the problem of matching both the luminosity
function and the Tully-Fisher relation.
Traditional N-body simulations suffer from a similar overmerging
problem (e.g.\ White 1976), which is of a purely numerical nature,
caused by two-body heating in dense environments (Carlberg 1994;
van Kampen 1995) when the mass resolution is too low. This is why
the statistical properties of EPS haloes and haloes found in N-body
simulations still agree quite well (e.g.\ Efstathiou et al.\ 1988;
Somerville et al.\ 1998 and references therein), despite the overmerging.

In order to circumvent these problems,
we use an N-body simulation technique that includes a built-in
recipe for galaxy halo formation, designed to prevent overmerging
(van Kampen 1995, 1997), to generate the halo population and its
formation and merger history. We show that this alone significantly changes
the predictions for the Tully-Fisher relation and the luminosity function.
In fact, it resolves most of the discrepancy sketched above, {\it and}\ allows
us to make the modelling more realistic by adding chemical evolution
and a merger-driven bursting mode of star formation to the modelling.

Star formation and its feedback to the interstellar medium are the least
understood ingredients of theories of galaxy formation. Usually one
considers a single mode of star formation, being either a continuous but
decaying rate of star formation from gas that cools within dark
haloes, or a discontinuous, bursting mode of star formation where stars
are formed at a high rate in short periods.
Here we will consider both, where the continuous mode of star formation
encompasses the case of many short bursts of star formation. We
reserve the term `starburst' for single major starburst events, lasting on
the order of 0.1 Gyr, as observed in luminous
infrared galaxies, for example (Sanders \& Mirabel 1996).

Once stars are formed, we apply the stellar population synthesis models
of Jimenez et al.\ (1998) to follow their evolution. We have enhanced
these models with a model for the evolution of the average metallicity
of the population, which depends on the starting metallicity.
Feedback to the surrounding material means that cooling properties of
that material will change with time, affecting the star formation rate,
and thus various other properties of the parent galaxy.

Several of our ingredients are thus different from earlier work:
we use numerical simulations for the formation and evolution
of the galaxy haloes, and we use newer population synthesis models,
which include chemical evolution. Nevertheless, it is still
useful to start from a model that is as similar as possible to a
published one. This allows us to assess the changes
in the model results due to the differences in modelling. 
We chose use the model of Cole et al.\ (1994) as a starting point.

The lay-out of this paper is as follows: we first discuss the main problems
in Section 2, focusing on the overmerging of galaxy haloes, and ways
of resolving the problems. In Section 3 we generate
distributions of galaxies for which we calculate luminosity functions and
the Tully-Fisher relation. We compare these to those found in previous
work, and to observed ones. We discuss the interpretation of the results,
pitfalls, and possible extensions, in the final section.

\newcount\japcount
\japcount=0
\def\japitem{\advance\japcount by 1
\smallskip\noindent\rlap{(\the\japcount)}\hglue\parindent\hangindent\parindent}

\section{The phenomenological model for galaxy formation}

\tx In this section, we summarize how
galaxy formation is modelled in this work. In many
respects, we deliberately follow the approach of other
groups, so as to aid comparison of results.
Nevertheless, our methods differ in some important
respects, as outlined below.
A detailed discussion of the modelling is given
in Appendix A; this section is intended as an
overview for non-specialists.

\subsection{Overview}

\tx The key ingredients of the model are:

\japitem
{\it The merging history of dark-matter haloes.}
This is often treated by Monte-Carlo realizations of the
analytic `extended Press-Schechter' formalism. That formalism
is a successful description of collapsed systems
with densities of order 200 times the mean, and of the
history of merging that produced such systems.
However, the theory ignores substructure:
a cluster-scale halo is treated as a single system.
The alternative is to measure halo substructure directly
via N-body simulation. Special techniques are used in this
paper to prevent galaxy-scale haloes undergoing `overmerging'
owing to inadequate numerical resolution.
By allowing substructure to survive, the merger
history is simplified and the number of halo mergers is
reduced.

\japitem
{\it The merging of galaxies within dark-matter haloes.}
Each halo contains a single galaxy at formation. When haloes 
merge, a criterion based on dynamical friction is used to decide how
many galaxies exist in the newly merged halo. The most massive
of those galaxies becomes the single central galaxy to which
gas can cool, while the others become its satellites. This
approximate argument is one of the more uncertain parts of
phenomenological galaxy formation schemes. We still use the
dynamical friction method, but our treatment of halo overmerging
means that part of the galaxy merging process is treated directly.
This should make our results more robust.

\japitem
{\it The history of gas within dark-matter haloes.}
When a halo first forms, it is assumed to have
an isothermal-sphere density profile. A fraction
$\Omega_b/\Omega$ of this is in the form of gas
at the virial temperature, which can cool to form
stars within a single galaxy at the centre of the halo.
Application of the standard radiative cooling
curve shows the rate at which this hot gas cools
below $10^4$~K, and is able to form stars.
Energy output from supernovae reheats some of the
cooled gas back to the hot phase. When haloes
merge, all hot gas is stripped and ends up in the new halo.
Thus, each halo maintains an internal account of the
amounts of gas being transferred between the two
phases, and consumed by the formation of stars.

\japitem
{\it Quiescent star formation.}
This is one of the more difficult areas to model.
Most authors assume a star-formation timescale that is
linearly proportional to the circular velocity
of the dark-matter halo in which the galaxy sits.
The star formation rate is equal the ratio of the amount of
cold gas available and the star-formation timescale.
The amount of cold gas available depends on the merger
history of the halo, the star formation history, and
the how much cold gas has been reheated by feedback
processes (discussed below).

\japitem
{\it Starbursts.}
We also introduce the idea that
the star-formation rate may suffer a sharp spike
following a major merger event. This is motivated
empirically by the existence of ultra-luminous
IRAS galaxies, and it allows hierarchical models
to yield behaviour resembling traditional
monolithic collapse. Our specific starburst model
is described in Appendix B.

\japitem
{\it Feedback from star-formation.}
The energy released from young stars heats cold
gas in proportion to the amount of star-formation,
returning it to the reservoir of hot gas. This
ingredient is essential to keep low-mass galaxies
and proto-galaxies from using up all their gas at
high redshift.

\table{1}{D}{\bf Table 1. \rm Parameters for the phenomenological
galaxy formation models discussed in Section 2, 3, and Appendix A.}
{\halign{%
\hskip4pt#\hfil&\hfil#\hskip4pt&\hfil#\hskip7pt&\hfil#\hskip4pt&\hfil
#\hskip4pt&\hfil#\hskip4pt&\hfil#\hskip4pt&\hfil#\hskip4pt&\hfil#\hskip2pt \cr
\multispan{9}\hrulefill\cr
\noalign{\vfilneg\vskip -0.4cm}
\multispan{9}\hrulefill\cr
\noalign{\smallskip}
Description & parameter & CAFNZ & m & n & c & s & a & b \cr
\noalign{\smallskip}
\multispan{9}\hrulefill\cr
\noalign{\smallskip}
Mass-to-light ratio: & \multispan{8} \cr
\ \ cosmological baryon fraction & $\Omega_{\rm b}$  & 
   0.06 &  0.06 & 0.06 & 0.06 &  0.06 & 0.06 & 0.06  \cr
\ \ $M({\rm all\ stars}) / M(m>0.1M_\odot)$ & $\Upsilon$ & 
    2.7 & 1.0 & 1.0 & 1.0 & 1.0  &  1.0 &  1.0  \cr
Continuous star formation: & \multispan{8} \cr
\ \ basic star formation time-scale & $\tau^0_*$ &  
     2  &    2   &    2  &    2  &  $10^6$ &  2 &    2  \cr
\ \ power-law index   & $\alpha^*$  &
    $-1.5$ & $-1.5$   & $-1.5$  & $-1.5$ & $-1.5$ & $-1.5$ & $-0.5$  \cr
Bursting star formation: & \multispan{8} \cr
\  \ basic star formation time-scale & $\tau^0_{\rm b}$ & 
    -  &    -   &  -    &  -  &  0.01 & 0.01 & 0.01  \cr
\  \ burst factor    & $f_{\rm b}$ &  
   -  &    -   &  -    &  -  &  $10^6$ &  100 &  100  \cr
Feedback (reheating of cooled gas by supernovae): & \multispan{8} \cr
\ \ feedback parameter      & $f_{\rm v}$ & 
   0.2 &  0.2   &  0.2  &  0.2  &  0.2  &  0.2 &  0.04 \cr
\ \ feedback normalization  & $V_{\rm hot}$ &
    140  &  140   &  140  &  140  &  140 &  140 &  100  \cr
\ \ feedback power-law index & $\alpha_{\rm hot}$ & 
   5.5 &  5.5   &  5.5  &  5.5  &  5.5 &  5.5 &  4.0  \cr 
Galaxy merging within haloes: & \multispan{8} \cr
\ \ dynamical friction time-scale & $\tau^0_{\rm mrg}$ &
  0.5$\tau_{\rm dyn}$ & 0.5$\tau_{\rm dyn}$ & 0.5$\tau_{\rm dyn}$ &
  0.5$\tau_{\rm dyn}$ & 0.5$\tau_{\rm dyn}$ & 0.5$\tau_{\rm dyn}$ &
  0.5$\tau_{\rm dyn}$ \cr
\ \ dynamical friction scaling law    & $\alpha_{\rm mrg}$ & 
  0.25  &  0.25 & 0.25 & 0.25 & 0.25 & 0.25 & 0.25  \cr
Stellar populations: & \multispan{8} \cr
\ \ initial mass function  & IMF  &
    Scalo & Salpeter  &  Salpeter  &  Salpeter  &  Salpeter  &
    Salpeter  &  Salpeter \cr
\ \ metallicity    & $Z$ &
   solar & solar & solar & evolving & evolving & evolving & evolving \cr
\noalign{\smallskip}
\multispan{9}\hrulefill\cr
\noalign{\vfilneg\vskip -0.4cm}
\multispan{9}\hrulefill\cr
\noalign{\smallskip}
}}

\japitem 
{\it Stellar evolution and populations.}
Having formed stars, we wish to predict the appearance
of the galaxy that results. For this, it is necessary to
assume an IMF, and to have a spectral synthesis code.
Our work assumes the spectral models of Jimenez et al.\ (1998);
for solar metallicity, the results are not greatly
different from those of other workers.
The IMF is generally taken to be Salpeter, but any choice is possible.
Unlike other workers, we take it as established that the
population of brown dwarfs makes a negligible contribution
to the total stellar mass density, and we do not
allow an adjustable $M/L$ ratio, $\Upsilon$, for the stellar
population.

\japitem
{\it Chemical evolution.}
It would of course be unrealistic to assume that
the entire star-formation history of a galaxy can unfold
at constant metallicity. The evolution of the metals
must be followed, for two reasons: (i) the cooling of the
hot gas depends on metal content; (ii) for a given age,
a stellar population of high metallicity will be much
redder. The models of Jimenez et al.\ (1998) allow
synthetic stellar populations of any metallicity to be
constructed. Appendix C discusses how the chemical
evolution of the gas is followed.

\subsection{Model parameters}

\tx This framework is rather general, and includes a number of
parameters. Note that there are also parameters involved in the
cosmological model. In this paper, we consider only
a single variant of CDM ($\Omega=1$, $h=0.5$, $n=1$,
$\Omega_b=0.06$, $\sigma_8=0.62$), in order to illustrate
the effects of changes in the galaxy-formation model.
This cosmological model is identical to the one used by
Cole et al.\ (1994).

The equations containing the specific parameterizations of
the above assumptions are given in Appedix A.
The values assumed for these parameters in the models
investigated in this paper are given in Table 1.

\figps{1}{S}{\psfig{file=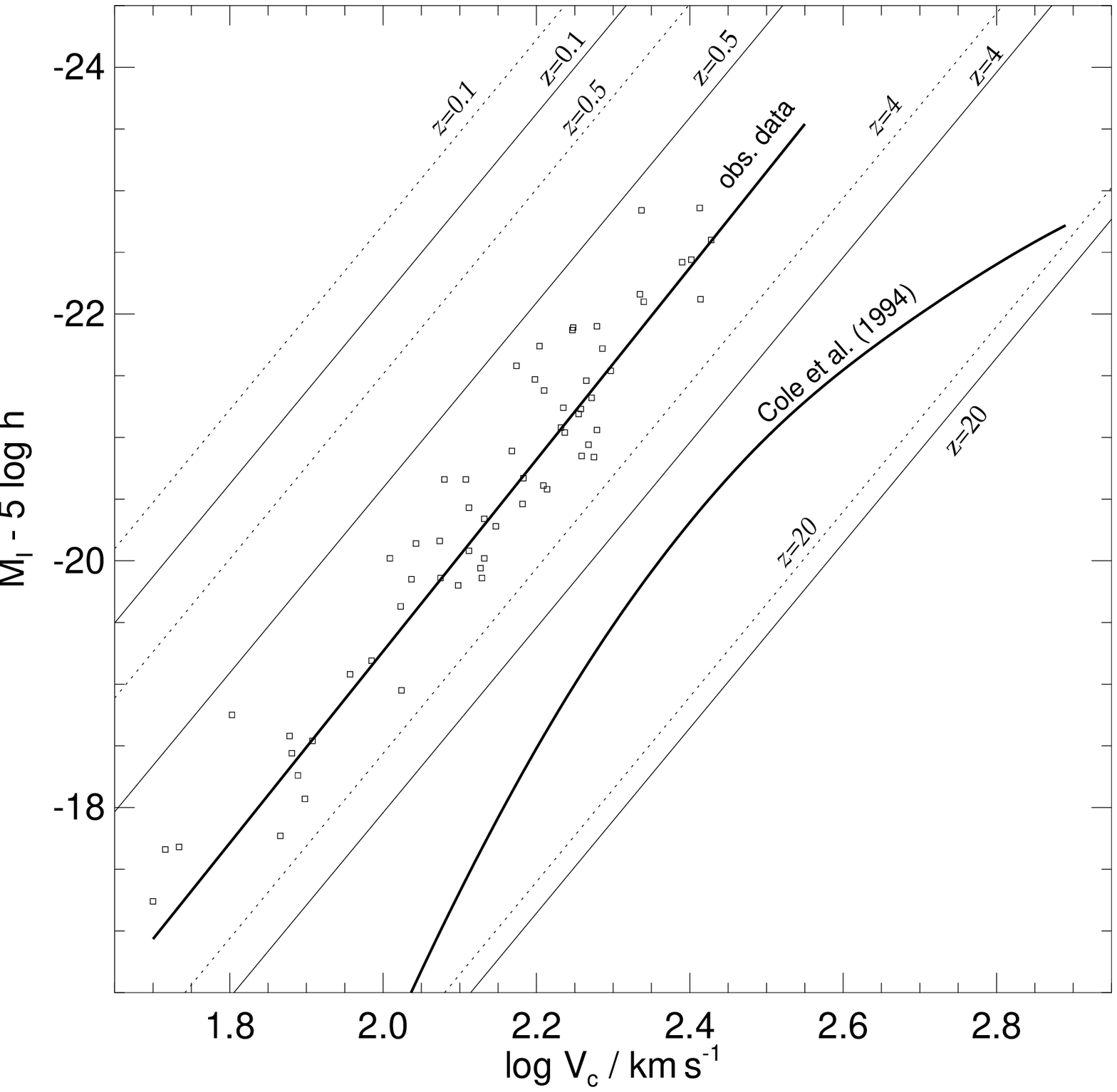,width=8.5cm,silent=}}
{{\bf Figure 1.} The Tully-Fisher relation: observational data and a
simple model: single redshift collapse, with all available gas forming
stars in an 0.1 Gyr burst (solid lines), or continuously with an
exponential decay-time of 10 Gyr (dotted lines).
The squares are individual galaxies as observed by Tully et al.\ (1998),
whereas the thick straight line is an average over four fits to
observational data by four different authors (see text for details).
The thick curved line is a fit to the model of Cole et al.\
(1994, curve taken from Heyl et al.\ 1995). Note that the models of
CAFNZ do have significant star formation up to the present epoch.}

\section{The Tully-Fisher / luminosity function discrepancy}

\tx Matching both the Tully-Fisher relation, which relates the infrared
rest-frame magnitude to the circular velocity $V_{\rm c}$ of the halo,
and the B-band luminosity function has been a problem for most published
phenomenological galaxy formation scenarios. In the following we use,
as a starting point, the models of Cole et al.\ (1994, CAFNZ from here on),
which show this discrepancy as well. When CAFNZ set their free parameters
to match the B-band luminosity function, they did not find a good
match to the I-band Tully-Fisher relation: it is either a
factor of two too large in $V_{\rm c}$, or 2-3 magnitudes too faint in I.

We believe that this is largely due to an excess of bright galaxies with
rather large circular velocities, caused by overmerging within
the (extended) Press-Schechter formalism used by CAFNZ. Because of
the overabundance of massive galaxies, CAFNZ were forced to put in place
several `brakes' in order to prevent the luminosity evolution from
overshooting.
Firstly, they adopt a value for the star-formation timescale $\tau_*^0$
which is half that of the value of 4 Gyr that is found in the
numerical simulations of
Navarro \& White (1993), on which CAFNZ based their star formation
models, and which was used to set many other parameters.
Secondly, they adopted a very large mass-to-light ratio for the
stellar population: they set $\Upsilon=2.7$, where $\Upsilon$ is defined
as the ratio of all stars to that of stars above 0.1 $M_\odot$.
Finally, they assume that all hot gas has
primordial metallicity (even gas reheated by solar metallicity supernovae),
which means it cools less rapidly than enriched gas and results in less
stars being formed per unit mass. They also assumed instant enrichment
to solar metallicity for their stellar populations, which helps keep
the B-band magnitudes faint.

\subsection{Building blocks for the Tully-Fisher relation}

\tx In order to provide a frame of reference for interpreting and
understanding the Tully-Fisher relation produced by our models,
we first greatly simplify the modelling by only incorporating some
of the ingredients. The very simplest model is one in which a galaxy
turns all its gas into stars at a single epoch of formation, either
in a short burst of star formation, or as an exponentionally decaying,
but continuous star formation mode. We will refer to this type of
model as a `single redshift star formation model'. The circular velocity
is calculated using the spherical collapse model (e.g.\ Peebles 1980),
where the density at first collapse is 178 times the background value,
which gives $L \propto M_{\rm gas} \propto V_{\rm c}^3 (1+z_{\rm form})^{3/2}$.

In Fig.\ 1 we plot the resulting Tully-Fisher relation for such a model,
where the single formation redshift (for all galaxies) is annotated for
each of the model relations. The straight thick solid line is an average
of fits to four datasets (Pierce \& Tully 1992; Mould et al.\ 1993 and
references therein; Giovanelli et al.\ 1997; Mathewson, Ford \& Buchhorn
1992), whereas the squares show the data of Young (1994),
in order to give an indication of the typical spread around the
observed relation.
This average relation corresponds to $L\propto V_{\rm c}^{3.1}$,
so very close to that predicted by the spherical collapse model.
The most obvious lesson to learn from this figure is that
we can reproduce the Tully-Fisher relation simply by forming all galaxies
between a redshift of 0.5 and 4, with a peak near $z=1-2$, and turning
all available gas into stars shortly after the formation event.
Galaxies will move straight down in the diagram if only part of the gas
is used up, so late star formation from a fraction of the gas produces
the same relation as turning all gas into stars at early times.

\subsection{Observed luminosity functions}

\tx In the models of CAFNZ the B and K-band luminosity function
(LF for short) were used to set some of their free parameters,
so that they matched well to the
observed luminosity functions known at the time: the B-band LF of Loveday
et al.\ (1992) and the K-band LF of Mobasher et al.\ (1986).
Present day datasets show quite a change in the observed LF:
Zucca et al.\ (1997) observe a much steeper faint end for the B-band LF,
while Gardner et al.\ (1997) and Glazebrook et al.\ (1995) observe a
K-band LF that is fainter at the bright end, although there is some
disagreement at the faint end. It must be gratifying to CAFNZ that their models
are closer to these new results that to the data of Loveday et al.\ (1992) and
Mobasher et al.\ (1986). The significant discrepancy with the Tully-Fisher
relation remains, however, and we attempt to resolve that in the remainder
of this Section.

\subsection{The galaxy halo population}

\subsubsection{Defining haloes}

\tx In most phenomenological galaxy formation recipes published
so far the galaxy halo population and its formation and merger
history are obtained from the extended Press-Schechter (EPS)
formalism (see the introduction for references and discussion).
It is essential to realize that the EPS formalism deals only
with systems of density approximately 200 times the mean,
which are {\it assumed}\ to be virialized.
Such sets of particles are readily
identified in N-body simulations e.g. by linking particles
via the percolation (a.k.a.\ `friends-of-friends') algorithm.
Indeed, the term `halo' is not infrequently taken implicitly to
denote only systems of this sort.
However, we feel this is an unfortunate useage, and we will use
the term in this paper in a more general sense: as a virialized
local density maximum. A rich cluster is an EPS halo, but 
the galaxies it contains are defined by dark-matter haloes
that constitute substructure in the cluster dark matter.
These systems collapsed before the cluster, and so have
present-day density contrasts well above 200. As the cluster forms,
the outer parts of these galaxy-scale haloes merge and
lose their identity; the high-density cores of the haloes will
nevertheless survive and remain virialized, and mark the locations
of galaxies within the cluster.
The aim of a galaxy formation model is to predict the
stellar content of these `embedded' galaxy-scale haloes.

\subsubsection{Overmerging}

\tx If one takes the galaxy haloes merger history from the EPS
formalism, all embedding haloes are retained, but embedded haloes are lost;
EPS records only when haloes are incorporated into a larger system,
but subsequent evolution within the new system is not followed.
We adopt the term `overmerging' to describe this
loss of information concerning the substructure in embedding haloes.

The use of the EPS formalism is usually motivated by the fact
that the distribution of masses of the haloes and their formation
histories are in good agreement with those found in N-body simulations
(Efstathiou et al.\ 1988; Somerville et al.\ 1998 and references therein).
However, this result depends on the choice of the group finder adopted
to identify haloes in the simulations (Suginohara \& Suto 1992).
The usual choice is percolation with a fixed linking-length, which
shares with the EPS formalism the property that embedded haloes
are not identified. It also identifies {\it any} overdense group,
not just virialized haloes.

Furthermore, overmerging also occurs in the N-body simulations to
which the EPS formalism is compared.
Small groups of particles that represent galaxy haloes
get disrupted by numerical two-body heating, especially
in large overdense systems (Carlberg 1994; van Kampen 1995).
If the numerical resolution is good enough, this problem is resolved
(Moore et al.\ 1998; Tormen, Diaferio \& Syer 1998),
but both the spatial and mass resolution need to be rather high:
Klypin, Gottl\"ober \& Kravtsov (1997) find that haloes survive for
resolutions of a few kpc and 10$^8$-10$^9$ $M_\odot$ respectively.
Unfortunately, for N-body simulations on a cosmological scale, this
requires the use of far too many particles (on the order of $10^9$)
to be practical, so a special N-body technique is needed instead.

\subsubsection{Resolving the overmerging problem}

\tx In order to prevent overmerging, we adopt a simulation technique
in which a galaxy halo formation recipe is added to an otherwise
standard N-body technique (van Kampen 1995, 1997).
The idea is that a group of particles that
has collapsed into a virialised system with a mass appropriate of a 
galactic halo is replaced by a single `halo particle'. Local
density percolation, also called adaptive friends-of-friends,
is adopted for finding the groups (van Kampen 1997).
This is designed to identify the embedded haloes that the 
traditional percolation group finder links up with their parent halo.
A Gaussian filter, with a length equal to the average nearest neighbour
distance for a Poisson particle distribution, is applied to calculate
a local density $n_{\rm G}$. This local density is then used to
calculate a local linking length for the percolation algorithm,
scaling as $n_{\rm G}^{-1/3}$. The linking
length also scales with the particle mass $m$ as $m^{1/2}$, in order
to account for the larger gravitational pull of the more massive halo
particles. The basic linking length (for mean density and non-halo
particles) is half the Poissonian average nearest neighbour distance.
 
The virial equilibrium criterion 
is put into a simplified form using the half-mass radius $R_{\rm h}$,
motivated by Spitzer (1969) who found that for many equilibrium systems
\newcount\virialB
\virialB=\eqnumber
$$\langle v^2\rangle \approx 0.4 {GM\over R_{\rm h}}\ . \eqno(\neweq)$$
A group is considered to be virialized when this criterion is satisfied
with a margin of 25 per cent.
The new halo particle is softened according to the size of the group
it replaces (thus conserving energy), and gets the position and momentum
of the centre-of-mass of the original group. Because groups of particles
can contain halo particles, merging is naturally included in this recipe as
well. Note that we do {\it not}\ assume that each halo particle hosts a
single galaxy. At formation, each halo particle may contain several
pre-existing galaxies, and the dynamical friction argument (ingredient
2 of Section 2.1) has to be invoked to see whether they merge. However,
our procedure clearly treats a good part of the general merging of
galaxies directly, which increases the robustness of the results.

If groups are sufficiently massive compared to the numerical resolution of
the code, they will avoid spurious numerical disruption in future evolution
of the density field. There is therefore no need to replace these systems
by a single particle. In fact, doing so would pose numerical problems as
supermassive simulation particles will interact violently with `normal'
N-body particles.
Therefore, an upper limit should be set to the mass of a halo particle.

However, the local density percolation group-finder we adopt here has been
designed to identify haloes in the field as well as in overdense regions.
This means that a smooth cluster containing no subhaloes will be identified
as one massive halo, while a cluster many subhaloes will not be identified
as a single halo. The centre of the cluster and the subhaloes will all
be identified as separate entities. This implies that there will
automatically be a maximum for the mass of the groups found,
depending on the environment. This is discussed in more detail in
van Kampen (1999, in preparation).

Preferrably the upper limit to the halo particle-mass should equal this
maximum group-mass, but as it depends on the group finder parameters,
it is hard to predict its value. The results should be robust
with respect to reasonable variations in this number, so the actual
choice is not too important. We set the upper mass limit to be
$7\times 10^{13} h^{-1}$ M$_\odot$, or about 1600 of the initial
simulation particles.

\figps{2}{D}{\psfig{file=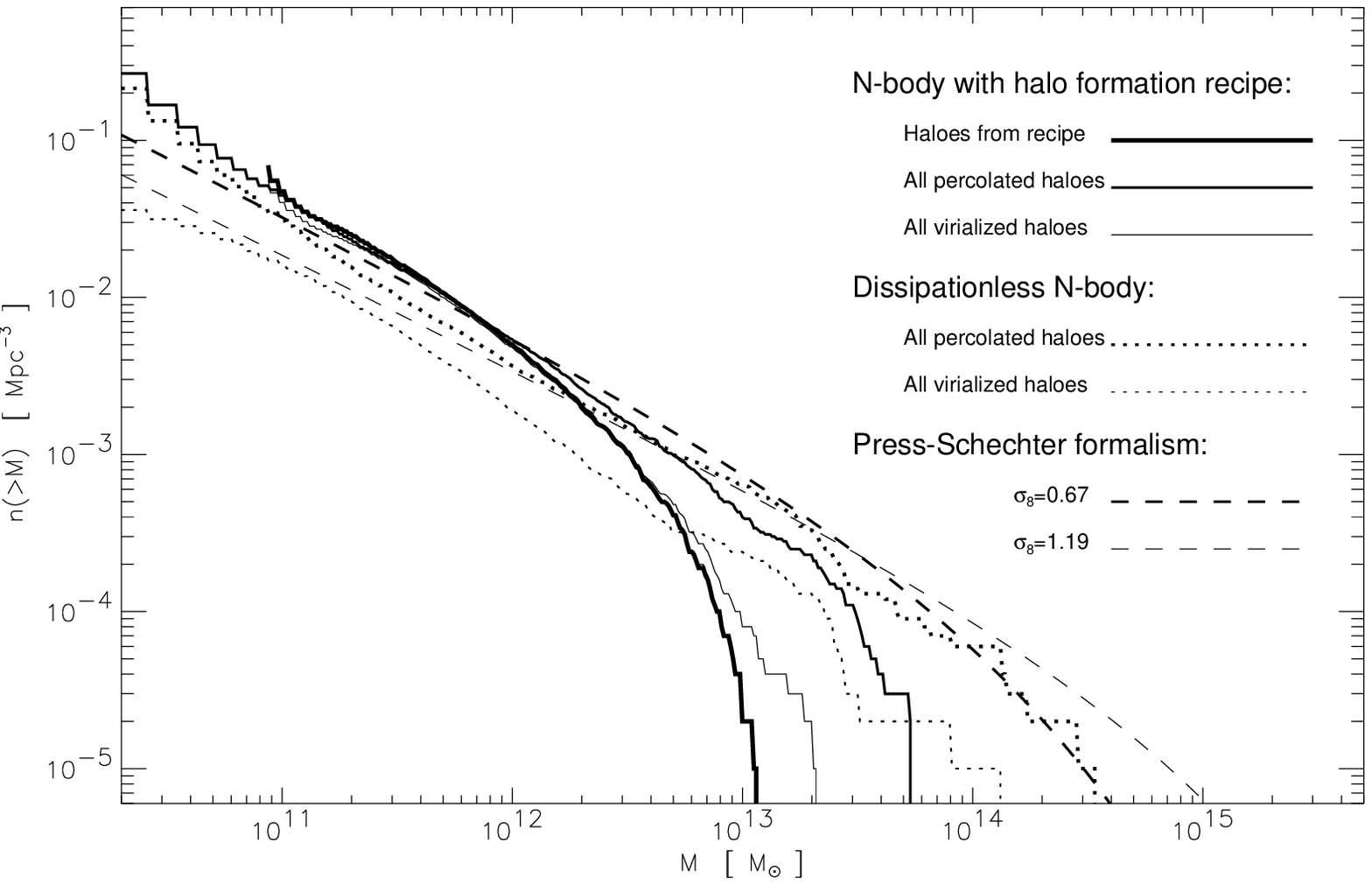,width=18.0cm,silent=}}
{{\bf Figure 2.} A comparison of galaxy halo cumulative mass distributions.
Dashed lines show the mass function for the Press-Schechter formalism for
standard CDM with two different normalizations: $\sigma_8=0.67$ (thick
dashed line), and the COBE normalization ($\sigma_8=1.19$, thin dashed line).
The thick dotted line represents the mass function for all haloes identified
in a standard dissipationless N-body simulation (for $\sigma_8=0.67$) using the
percolation algorithm, while the thin dotted line
shows the subset of those haloes that have virialized. The thickest solid line
shows the mass functions for haloes formed using the recipe of van Kampen
(1995, 1997), starting from the same initial conditions. Including all
haloes found from the percolation algorithm time at the final time
results in the mass function depicted by the thinnest solid line,
while the solid line with intermediate thickness shows the subset of
these haloes that have virialized. Note the difference between the thin solid
line and the thin dotted line, which were obtained using the same group
finder.}

\subsubsection{Our halo population}

\tx We performed two high-resolution simulations, using the methods
described above, with a (spherical) volume of $(23.2 h^{-1} {\rm Mpc})^3$,
using around $8\times10^5$ particles.
The first simulation started from initial conditions contrained
to form a galaxy cluster in the centre of the simulation volume,
while the second simulation started from unconstrained initial
conditions. We will refer to the first simulation as the
`cluster model', and to the second one as the `field model'.

The importance of overmerging is most clearly illustrated by
looking at cumulative dark halo mass functions as produced by the
Press-Schechter formalism, a standard N-body simulation, and the
present method, as described above.

In Fig.\ 2 we compare all these to each other. The thick dashed line is
the halo function from the Press-Schechter formalism for the cosmology
adopted here, with $\sigma_8=0.62$. The {\it COBE}\ normalized version
(i.e.\ $\sigma_8=1.19$) is also shown, as a thin dashed line.
We ran the field model without the
galaxy halo formation recipe switched on, i.e.\ just as a traditional
N-body simulation, and obtained the mass function using the same
group finder (see Section 3.3.3) used in the full simulation technique.
This mass function
is plotted in Fig.\ 2 as a thick dotted line. It clearly matches the
Press-Schechter mass function. However, if we now plot the mass
function for {\it virialized haloes only}, being haloes which
satisfy eq.\ (1) within 25 per cent, it falls short by a factor
of two (thin dotted line). This is because in the EPS
formalism all collapsed haloes are assumed to virialize, while
haloes as identified in numerical simulations change through merging,
secondary infall, tidal forces, etc., and are therefore not able to
remain virialized at all times. This is even true for galaxy cluster
haloes (Natarajan, Hjorth \& van Kampen 1997).
Thus, virialized haloes form a subset of all haloes at any one time.
Indeed, the haloes formed in our simulations contain about 25
per cent of all the mass in the universe, while in the EPS formalism
all matter is contained in haloes, by design.

The halo functions that are produced by the full simulations are shown
in Fig.\ 2 as solid lines. The thickest of the three lines represents
the halo function for the galaxy haloes identified by the recipe.
The thinnest solid line shows the effect of adding the virialized haloes
identified at $z=0$, including the ones above the upper mass-limit. Note that
there are only a few more haloes added, so our upper mass-limit for halo
particles is close to the maximum mass found from the halo formation
recipe employed.

The right-most solid line shows the mass function for all percolated systems
at $z=0$: galaxy halo particles, newly virialized haloes, and all other
collapsed haloes. We see that locking matter in haloes at early times
has the effect that the largest mass haloes that form in both
the Press-Schechter formalism and standard N-body simulations are now
correctly resolved into subhaloes and are not counted as single
objects. More interestingly, the maximum mass found for virialized haloes is
close to that found for observed galaxies. The core of the simulated cluster
is one of these most massive virialized haloes. 

Because of our choice for the upper-limit to halo particles,
the most massive haloes are not treated as single particles. However,
they are identified in the simulation, and are incorporated into the
merger tree used for the phenomenological galaxy formation model.

\section{Resolving the Tully-Fisher / luminosity function discrepancy}

\tx Before showing our model results, a word of warning:
throughout this section we plot Tully-Fisher diagrams that contain
{\it all}\ model galaxies found from the modelling. We should therefore
bear in mind that only a subset of all these galaxies, i.e.\ the spirals,
actually belong in such a diagram.
As we do not model the morphologies of the galaxies, we aim to
at least cover the region of $V_{\rm c}-M_{\rm I}$ space occupied by the
observational data.
Galaxies outside that region may in fact be regarded as ellipticals, S0's,
and irregular galaxies, and their properties should be tested using
other observational relations, like the Faber-Jackson relation, which is
the analogue of the Tully-Fisher relation for ellipticals.

One could be tempted to merge the Tully-Fisher
and Faber-Jackson relations into one diagram, as the circular velocity
is related to the velocity dispersion through the assumption of
an isothermal density profile. However, the Faber-Jackson relation
concerns the {\it central}\ velocity dispersion, which is not equal to,
and usually larger than, the velocity dispersion of the dark matter halo.
White (1979) found that merger remnants tend towards a $r^{-3}$ density
profile, so if elliptical form through merging, they should have declining
circular velocity profiles. Observationally, it is very difficult to measure
halo properties of generally gas-poor elliptical galaxies, whereas spiral
galaxies have sufficient cold gas at large radii to act as a tracer of
the halo potential.
Thus, we need more detailed modelling before we can attempt a match to
the Faber-Jackson relation, and we therefore restrict ourselves to the
Tully-Fisher relation for now.

\subsection{Repeating CAFNZ: model {\rm m}}

\tx To see the effect of using a merger tree obtained from numerical
simulations that employ the halo formation recipe of van Kampen (1997),
instead of a merger tree generated according to the extended
Press-Schechter formalism, we applied the same prescription of gas
dynamics and star formation that CAFNZ used, with the same parameters
($\tau_*^0=2$ Gyr etc.). The only other difference with the CAFNZ model
beside the halo population is the choice of IMF and stellar population
models (see Table 1), but we believe these differences to be marginal
(see Appendix A). The spherical infall model has been adopted to
calculate the circular velocity, as in CAFNZ.

The result for the cluster simulation is plotted in Fig.\ 3a.
What is apparent from this figure is that we reproduce the CAFNZ
Tully-Fisher relation, but miss the high-$V_{\rm c}$ end of it.
Because our model galaxy haloes do not overmerge, the large-$V_{\rm c}$
haloes found by CAFNZ are divided into a number of smaller haloes
in our modelling (see Section 3.3).
The effect on the B and K band luminosity funcion is a shift to the fainter
end, as is shown in Fig.\ 4a. The thick line is the luminosity function
found by CAFNZ, while the symbols with error bars represent the
observational data described above. Our model {\rm m}
results quite clearly fit neither the observations nor the CAFNZ models.
But this is actually a very nice result, as it allows us to resolve much
of the Tully-Fisher / luminosity function discrepancy, as follows.

\figps{3}{D}{\psfig{file=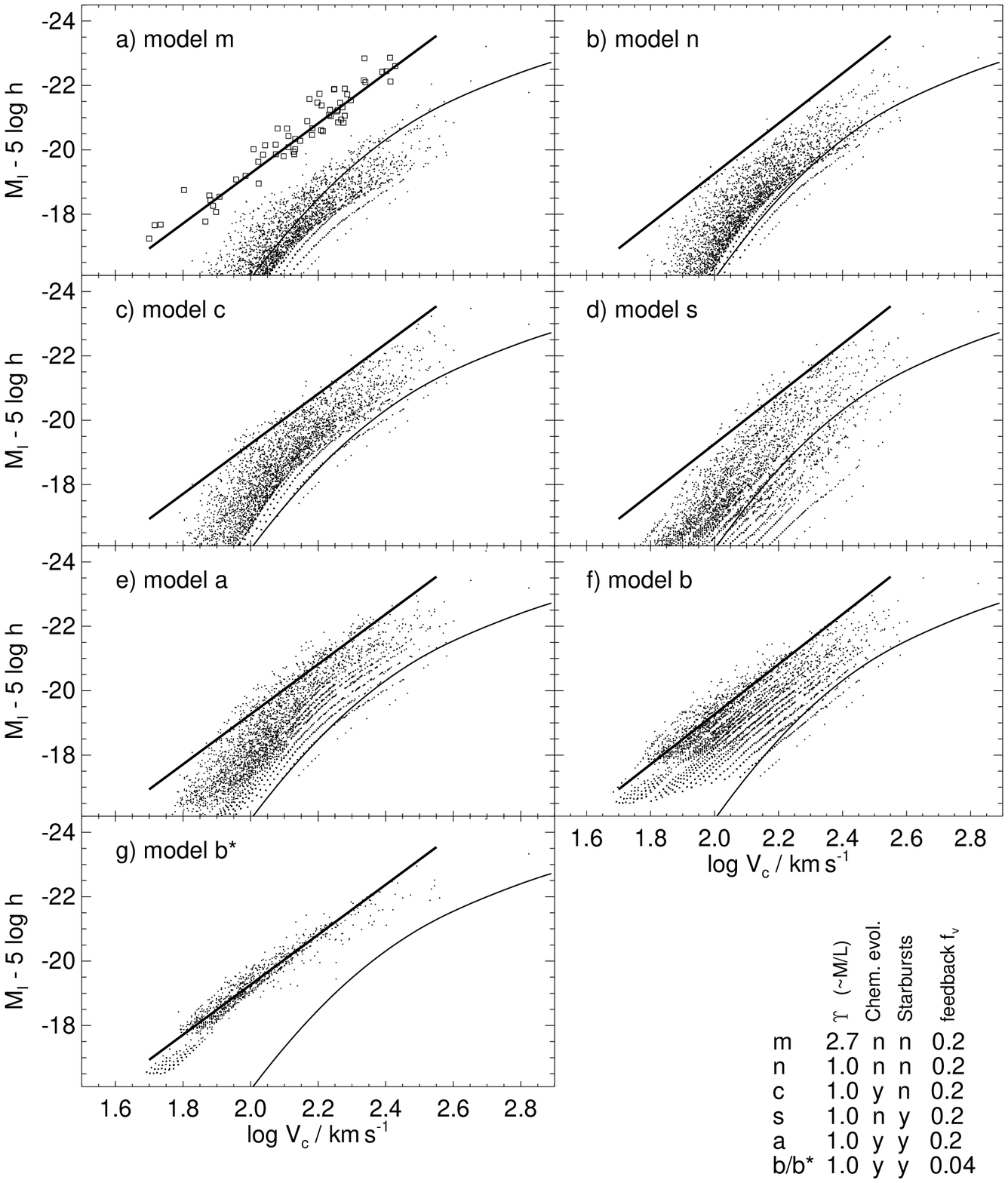,width=17.2cm,silent=}}
{{\bf Figure 3.} I-band Tully-Fisher relation from our
modelling (dots) for various choices of parameters. The first six panels
correspond to the six models discussed in the text, whereas the bottom
left panel shows model {\rm b} for a subset of the data that should
represent an observational dataset (see Section 3.4.6 for details).
We refer to the legenda and to Table 1 for the exact model specifications.
Also shown are fits to the distributions from CAFNZ (curved thin solid line),
observational data averaged over four different sets of data (straight
thick solid line, see main text for details). In the top left panel we
added observational data by Tully et al.\ (1998), shown as squares.
Note that the specific redshifts at which the galaxy formation recipe
is applied are clearly visible for model {\rm b}, and to a lesser extent
for the other models, due to the simplified modelling of starbursts. 
}

\figps{4}{D}{\vskip-0.5cm\psfig{file=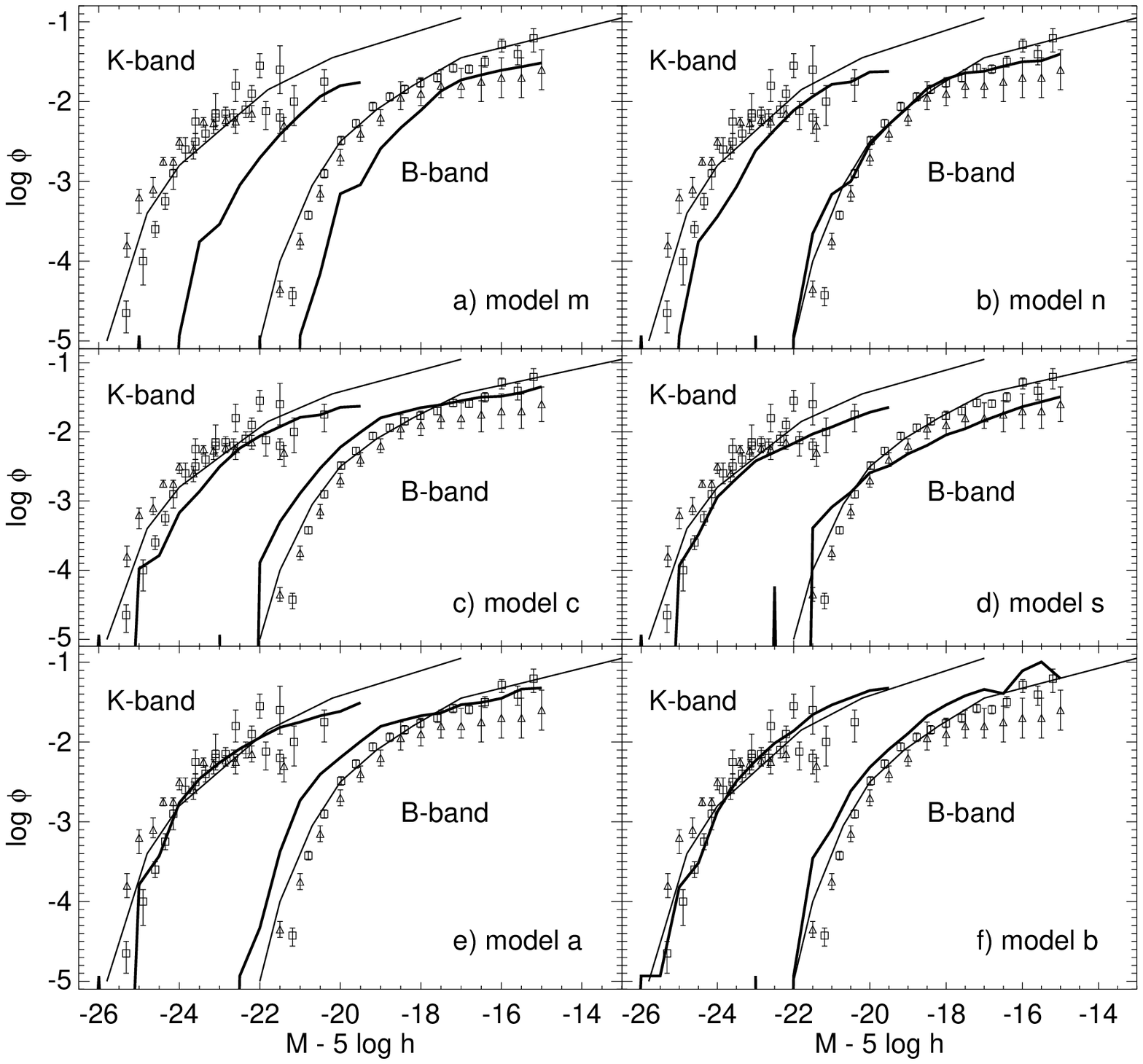,width=17.5cm,silent=}}
{\vskip 0.0cm{\bf Figure 4.} B and K band luminosity functions from our
modelling (thick solid lines), corresponding to the Tully-Fisher
relations shown in Fig.\ 5 (in the same order), and those from CAFNZ
(thin solid lines). Please refer to the legenda of Fig.\ 3 and Table 1
for details on the models. Also shown are
old observational data (triangles) from Loveday et al.\ (1992) for the
B-band and Mobasher et al.\ (1993) for the K-band, and new observational
data (squares) from Zucca et al.\ (1997) for the B-band and
Gardner et al.\ (1997) and Glazebrook et al.\ (1995) for the K-band.}

\subsection{Setting $\Upsilon=1$: model {\rm n}}

\tx Considering the `brakes' that CAFNZ had to apply to their
models, it should come as a relief that our model galaxies are too
faint in B and K, since the I-band magnitudes were already
too faint for the Tully-Fisher relation to fit. We can thus apply
an {\it overall brightening}\ to the modelling in order to simultaneously
match the Tully-Fisher relation and both luminosity functions. This is
easily achieved by setting $\Upsilon=1$ instead of $\Upsilon=2.7$,
which was the value that CAFNZ needed to adopt.
This brightens the I-band Tully-Fisher relation by just over a magnitude,
with the result that the predicted distribution of points now
overlaps with the observational data (Fig.\ 3b).
If we assume that the brightest galaxies in I (for a given $V_{\rm c}$)
are those that were still forming stars up to a fairly recent epoch,
and are therefore likely to be spirals,
the predicted Tully-Fisher relation can be considered encouraging,
except for the faint end.
We discuss solutions to that at the end of this Section.

The luminosity functions are shown in Fig.\ 4b, in the corresponding panel.
The luminosity function matches the observations very well in the B-band,
but falls short in the K-band. In the next section we therefore introduce
two more ingredients that not only render the modelling more realistic,
but provides a better to the K-band luminosity function:
chemical evolution and starbursts.

\section{Chemical evolution and starbursts}

\subsection{Adding chemical evolution: model {\rm c}}

\tx Instead of adopting primordial metallicity for the cooling
function, and constant solar metallicity for the stellar
populations, we now incorporate a chemical evolution model.
The details of the method we use is described
in Appendix C. This means that we use a different set of 
stellar population models, with evolving metallicities.
Once a population forms, the {\it initial}\ metallicity and the star
formation timescale $\tau_*$ determines what the metallicity will be at
later times. The enrichment of hot gas by chemically evolving stellar
populations will increase the number of stars forming at later times,
because the cooling function depends on the metallicity. We assume
enrichment to be efficient, i.e.\ the metallicity of the cooling gas
is equal to that of the populations it cools to (mass-averaged over
co-existing populations in merged galaxies).

The effect on the Tully-Fisher relation of the inclusion of chemical
evolution can be seen in Fig.\ 3c, whereas the effect on the
luminosity functions is shown in Fig.\ 4c. We see that the B-band
luminosity function ends up too bright, whereas the K-band
one is still too faint. In other words, most galaxies end up
too blue. The reason for this is that chemical enrichment boosts late
stars formation with relatively slowly decaying star formation rates.
This means that we have to find a way to form stars earlier and with much
more rapidly decaying star formation rates. This leads us to the conclusion
that we need to incoorporate a bursting mode of star formation, driven
by merging activity, which is most significant at early times (e.g.\
Carlberg 1990).

Another possibility is the addition of a dust model, which will
typically render galaxies redder. However, we leave this for
future work to explore.

\subsection{Models with starbursts only: model {\rm s}}

\tx In order to see the effect of adding a bursting mode of star formation,
we first look at a burst-only model in which continuous galaxy formation
is effectively switched off. We make the burst as strong as possible,
with $f_{\rm b}$ set to a very large value, and with a very short decay-time.
However, the duration of the starburst is set as described in Appendix B.

The Tully-Fisher relation produced by this pure starburst model is shown
Fig.\ 3d. It is quite similar to the chemical evolution model, in the
sense that more stars are formed, which brings enough galaxies on or
towards the Tully-Fisher relation.
The spread is a bit larger, as the earliest galaxies fade more rapidly if
they experience just a single starburst, therefore ending up fainter than
in model {\rm c}, whereas the late mergers are brighter because more
gas is used up at the present epoch than in model {\rm c}, which has a
much larger star formation time-scale.

However, the luminosity functions, as shown in Fig.\ 4d, are different
from the chemical evolution ones in the sense that the colours end up much
redder, providing a better match to the observational data for both bands.

\subsection{Models with both chemical evolution and starbursts: models
{\rm a} and {\rm b}}

\tx The pure starburst model produces fairly monochromatic
galaxies: all galaxies redden quickly and effectively stop forming stars,
because most mergers happen at high redshifts (e.g.\ Carlberg 1990), with a
merger rate peak at $z \approx 1$ (van Kampen 1997).
If we now switch on continuous galaxy formation again, those galaxies
that have some gas left (likely to be spirals), can form a (cosmologically)
young blue population that broadens the colour distribution {\it and}\
moves those galaxies on top of the observed Tully-Fisher relation. After all,
this relation is based on spiral galaxies in which some star formation
activity is typically found. In order to achieve this we set
$f_{\rm b}=100$ and $\tau^0_{*,\rm b}=0.01$ Gyr.

The resulting Tully-Fisher relation is plotted in Fig.\ 3e, and the
luminosity functions in Fig.\ 4e. They all match the observational data
reasonably well, except for the faint end of the Tully-Fisher relation.
However, we can easily `lift' this faint end by decreasing the amount
of feedback to $f_{\rm v}=0.04$, which comprises model {\rm b}.
The resulting Tully-Fisher relation is plotted in Fig.\ 3f, and this
is clearly shown to work. It slightly improves the B-band luminosity
function at the bright end, but at the same time worsens it at the
faint end, as shown in Figs.\ 4f. 

Finally, note that the remaining deviations from the observed luminosity
functions are likely to be resolved by taking into account dust, which
will redden the colours of the galaxies.


\subsubsection{Observational selection}

\tx So far we have made no attempt at discriminating between the model
galaxies, which makes quite a difference for the Tully-Fisher relation, as
only spiral galaxies enter the relation, and only those for which a reliable
circular velocity can be measured. This means that spiral galaxies showing
any sign of interaction are typically excluded from an observational sample.
Thus, we should disregard any model galaxy that underwent a major merging
event within the last 1-2 Gyr. Furthermore, galaxies that have not been
part of a major merger event for a long time, will have a much reduced
star formation rate, and are therefore not likely to possess an obvious
disk component. So we should also disregard galaxies that have formed at
fairly high redshift and evolved passively up to the present epoch.

We select an `observational Tully-Fisher sample' from all model galaxies
by selecting only those galaxies that satisfy
1 Gyr $< t_{\rm last} <$ 6 Gyr, where $t_{\rm last}$ is the lookback time
for the last major merging (or formation) event. This corresponds to
(approximately) a redshift range of $0.05<z_{\rm last}<0.5$.
These are plotted in Fig.\ 3g. Clearly, this subset of the model data
provides a reasonable match to the observational Tully-Fisher data.
However, this attempt at observational selection is rather crude,
and will need to be refined in future work. A proper disk model is a
necessity, as is a dust model which takes the inclination of the disk
into account.

%
%

\section{Discussion and conclusions}

\tx We have argued that overmerging is an important problem for
phenomenological galaxy formation models, and is largely responsible
for the discrepancy found in earlier work between the Tully-Fisher
relation and the luminosity function. We resolved the overmerging
problem by using an N-body simulation technique that includes a galaxy
halo formation recipe. We combined this technique with simplified gas
dynamics and star formation physics, and a recent stellar population
model, in order to phenomenologically describe galaxy formation and
evolution. We included chemical evolution, and two modes of star formation:
major bursts of star formation, and a quiescent mode in which stars
form more continuously.

With this set-up we match both the B and K band
luminosity function and the I-band Tully-Fisher relation, for an
$\Omega=1$ standard CDM structure formation scenario. Resolving the
overmerging problem is the major contributor to this result,
but the inclusion of chemical evolution and starbursts are also
important ingredients.

The new ingredients we have added to the modelling of galaxy formation
are needed in order to make the models more realistic, and are not
introduced simply in order to give yet more free parameters. Nevertheless,
our resolution to the Tully-Fisher / luminosity function discrepancy
may well not be unique, and various other changes to the ingredients of the
phenomenological galaxy formation recipe might produce similar
results. For example, we have not studied the influence cosmological
parameters have on the model galaxy populations, where $\Omega$,
$\Lambda$, and $\sigma_8$ are likely to be the important parameters.
Other types of ingredients are possible as well:
Somerville \& Primack (1998) resolve some of the discrepancy using
a dust extinction model plus a halo-disk approach to feedback.

We intend to explore these issues in future work. One way of resolving
the worries about degeneracies in the cosmological/physical parameter
space will be to include data at intermediate and high redshifts.
Nevertheless, although the present work has concentrated on modelling the
properties of the low-redshift universe only, we believe that the
issues we have raised are sufficiently general that they will invariably
be part of any successful model for galaxy formation.

\section*{Acknowledgements}

\tx We thank the anonymous referee for a report that encouraged us to clarify some
parts of this paper, and Carlton Baugh, Shaun Cole, and Eduard Thommes for
useful comments.

%
%

\section*{references}

\bibitem Barnes J.E., Hut P., 1986, Nat, 324, 446
\bibitem Binney J., 1977, ApJ, 215, 483
\bibitem Bond J.R., Cole S., Efstathiou G., Kaiser N., 1991, ApJ, 379, 440
\bibitem Bower R.J., 1991, MNRAS, 248, 332
\bibitem Carlberg R.G., 1990, ApJ, 350, 505
\bibitem Carlberg R.G., 1994, ApJ, 433, 468
\bibitem Cole S., Arag\'on-Salamanca A., Frenk C.S., Navarro J.F., Zepf S.E.,
1994, MNRAS, 271, 781
\bibitem Davis M., Efstathiou G., Frenk C.S., White S.D.M., 1985, ApJ, 292, 371
\bibitem Efstathiou G., Frenk C.S., White S.D.M., Davis M., 1988, MNRAS, 235, 715
\bibitem Elmegreen B., 1999, in Beckman J., ed., Formation and evolution of
galaxies on cosmological timescales, Cambridge University Press
\bibitem Faber S.M., Gallagher J.S., 1979, ARAA, 17, 135
\bibitem Gardner J.P., Sharples R.M., Frenk C.S., Carrasco B.E., 1997, ApJ, 480, 99
\bibitem Giovanelli R., Haynes M.P., da Costa L.N., Freudling W., Salzer J.J.,
Wegner G., ApJ, 477, L1
\bibitem Glazebrook K., Peacock J.A., Miller L., Collins C., 1995, MNRAS, 275, 169
\bibitem Heavens A.F., Jimenez R., 1999, submitted to MNRAS
\bibitem Heyl J.S., Cole S., Frenk C.S., Navarro J.F., 1995, MNRAS, 274, 755
\bibitem Jimenez R., Padoan P., Matteucci F., Heavens A., 1998, MNRAS, 299, 123
\bibitem Jimenez R., Dunlop J., Peacock J., MacDonald J., J{\o}rgenson U.G., 1999,
MNRAS, in press
\bibitem Kauffman G., White S.D.M., 1993, MNRAS, 261, 921
\bibitem Kauffman G., White S.D.M., Guiderdoni, 1993, MNRAS, 264, 201
\bibitem Kauffman G., Colberg J.M., Diaferio A., White S.D.M., 1998,
astro-ph/9805283
\bibitem Klypin A., Gottl\"ober S., Kravtsov A.V., 1997, astro-ph/9708191
\bibitem Kurucz R., 1992, ATLAS9 Stellar Atmosphere
Programs and 2 km/s Grid CDROM Vol. 13
\bibitem Lacey C.G., Cole S., 1993, MNRAS, 262, 627
\bibitem Longair M.S., 1998, Galaxy formation, Springer Verlag
\bibitem Loveday J., Peterson B.A., Efstathiou G., Maddox S.J., 1992,
ApJ, 90, 338
\bibitem Mathewson D.S., Ford V.L., Buchhorn M., 1992, ApJS, 81, 413
\bibitem Matteucci F., Francois P., 1989, MNRAS, 239, 885
\bibitem Mihos J.C., Hernquist L., 1994, ApJ, 431, L9
\bibitem Mihos J.C., Hernquist L., 1996, ApJ, 464, 641
\bibitem Mobasher B., Ellis R.S., Sharples R.M., 1986, MNRAS, 223, 11
\bibitem Moore B., Governato F., Quinn T., Stadel J., Lake G., 1998, ApJ, 499, L5
\bibitem Mould J.R., Akeson R.L. Bothun G.D., Han M-S., Huchra J.P.,
Roth J., Schommer R.A., 1993, ApJ, 409, 14
\bibitem Natarajan P., Hjorth J., van Kampen E., 1997, MNRAS, 286, 329
\bibitem Navarro J.F., White S.D.M., 1993, MNRAS, 265, 271
\bibitem Peebles P.J.M., 1980, Large-scale structure in the Universe, Princeton
\bibitem Pierce M.J., Tully R.B., 1992, ApJ, 387, 47
\bibitem Press W.H., Schechter P., 1974, ApJ, 187, 425
\bibitem Rees M.J., Ostriker J.P., 1977, MNRAS, 179, 541 
\bibitem Roberts M.S., Haynes M.P., 1994, ARAA, 32, 115
\bibitem Sanders D.B., Mirabel I.F., 1996, ARAA, 34, 749 
\bibitem Silk J., 1977, ApJ, 211, 638
\bibitem Somerville R.S., Lemson G., Kolatt T.S., Dekel A., 1998, astro-ph/9807277
\bibitem Somerville R.S., Primack J.R., 1998, astro-ph/9802268
\bibitem Spitzer Jr.\ L., 1969, ApJ, 158, L139
\bibitem Suginohara T., Suto Y., 1992, ApJ, 396, 395
\bibitem Sutherland R., Dopita M.A., 1993, ApJS, 88, 253
\bibitem Tormen G., Diaferio A., Syer D., 1998, 299, 728
\bibitem Tully R.B., Pierce M., Huang J.-S., Saunders W., Verheijen M.,
Witchalls P., 1998, AJ, 115, 2264
\bibitem van Kampen E., 1995, MNRAS, 273, 295
\bibitem van Kampen E., 1997, in Clarke D.A., West M.J., eds., Proc. 12th
`Kingston meeting' on Theoretical Astrophysics: Computational Astrophysics,
ASP Conf. Ser. Vol. 123. Astron. Soc. Pac., San Francisco, p. 231,
astro-ph/9904270
\bibitem White S.D.M., 1976, MNRAS, 177, 717
\bibitem White S.D.M., 1979, MNRAS, 189, 831
\bibitem White S.D.M., 1996, in Schaeffer et al., eds., Cosmology and
Large-scale structure, Proc. 60th Les Houches School, Elsevier, p.349
\bibitem Zucca E., Zamorani G., Vettolani G., Cappi A.S., Merighi R., Mignoli M.,
Stirpe G.M., MacGillivray H., et al., 1997, A\&A, 326, 477

%
%

\figps{5}{S}{\psfig{file=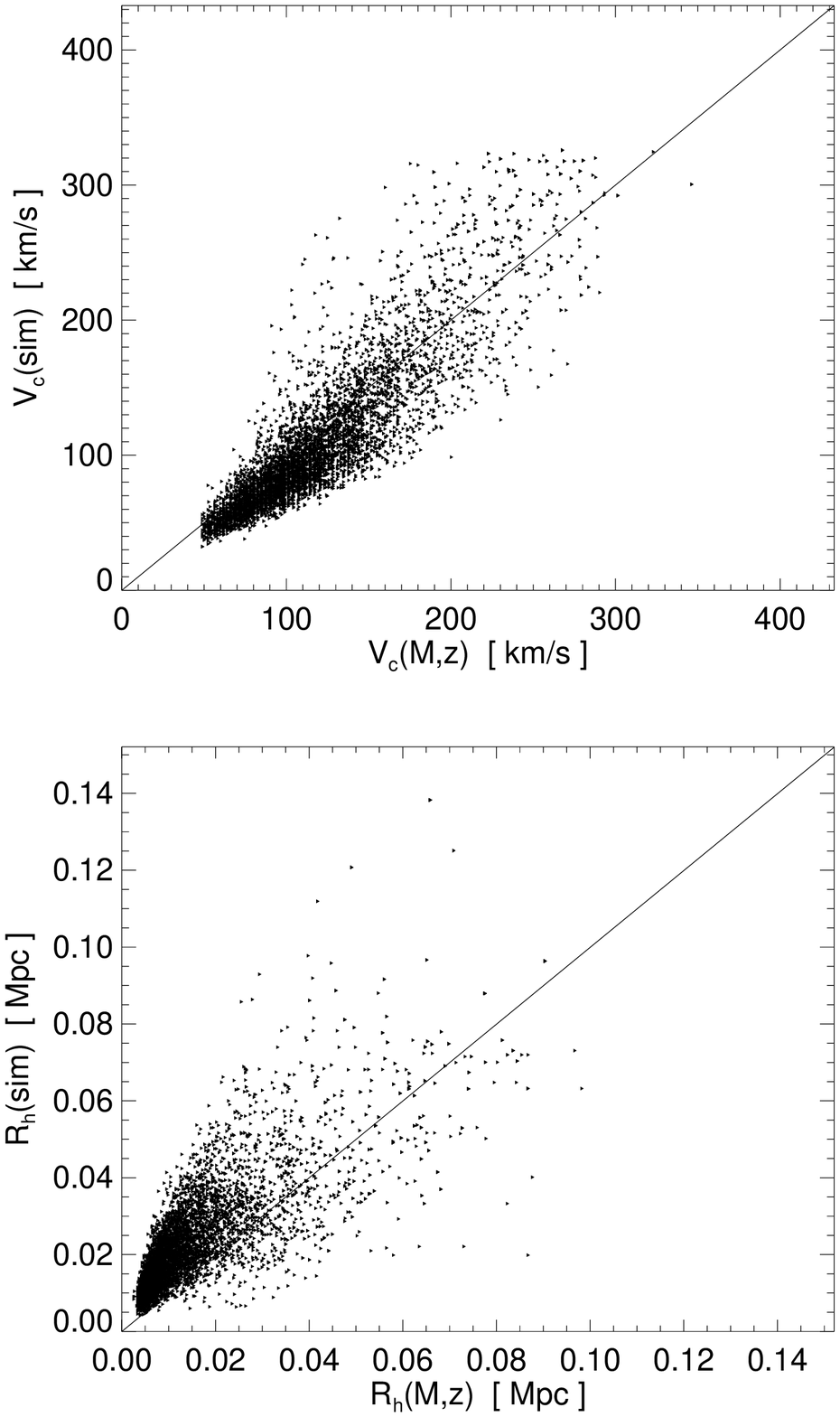,width=8.5cm,silent=}}
{{\bf Figure A1.} Comparison of circular velocity from the spherical
infall model and from N-body simulations (top panel), and a similar
comparison for the half-mass radii (bottom panel). The solid lines
simply indicate equality.}




\section*{Appendix A: details of the galaxy formation recipe}

\eqnumber=1
\tx This Appendix provides some further details of the phenomenological modelling
of galaxy formation adopted for this paper. We chose to use the modelling
of CAFNZ as a starting point, but note that many authors
use similar models with differerences in the details or in the choice
of parameters. Significant changes to the CAFNZ models are:
the halo merger history and mass function, the inclusion of star bursts,
and the modelling of chemical evolution.
All parameters encountered in the various ingredients of the model
are listed in Table 1, along with their values for the specific models
discussed in the main paper.

\section*{\it Halo population and its merger history}

\tx In most phenomenological galaxy formation recipes published
so far the galaxy halo population and its formation and merger
history are obtained from the extended Press-Schechter (EPS) model
(see the introduction for references). Instead, we use the haloes
found an N-body simulations which includes a recipe for the formation
of galaxy haloes (van Kampen 1995, 1997), as described in Section 3.3.3.
The actual N-body code used is the Barnes \& Hut (1986) treecode.

Haloes are assumed to form according to the spherical collapse model
(e.g.\ Peebles 1980), and are assumed to settle as isothermal spheres,
with a constant circular velocity $V_{\rm c}$.
The spherical collapse model provides a relation between the mass
of the halo, its circular velocity, and its formation redshift
(e.g.\ White 1996) :
$$V_{\rm c} = \Bigl({M_{\rm halo}\over 2.35\times 10^5 h^{-1}
        {\rm M}_\odot}\Bigr)^{1\over 3}
	\Bigl(1+z_{\rm form}\Bigr)^{1\over 2}\ 
	{\rm km s}^{-1}\ , \eqno(A\neweq)$$
where a truncated isothermal density profile has been assumed.
One is forced to adopt this crude model as the extended
Press-Schechter formalism that is typically used does not
provide information on the kinematics of the haloes.
The N-body simulations do provide us with that information, so we can
obtain $V_{\rm c}$ directly from the velocity dispersion of the halo,
which is in virial equilibrium by definition. Again assuming an
isothermal density profile, we have
$$V_{\rm c}=\Bigl({2\over 3}\Bigr)^{1\over 2} \langle v^2\rangle^{1\over 2} 
        \approx \Bigl(0.33 {GM_{\rm halo}\over R_{\rm h}}\Bigr)^{1\over 2},
        \eqno(A\neweq)$$
where $\langle v^2\rangle$ is obtained from the numerical simulation.
Basically, we allow dark haloes with the same mass and $z_{\rm form}$ to
have a range of values for $R_{\rm h}$, and therefore different $V_{\rm c}$,
while CAFNZ assume the fixed value
$$\Bigl({R_{\rm h}\over 100 h^{-1}{\rm kpc}}\Bigr) =
        \Bigl({M_{\rm halo}\over 6.37\times 10^{12} h^{-1} {\rm M}_\odot}
        \Bigr)^{1\over 3} {1\over 1+z_{\rm form}}. \eqno(A\neweq)$$

In order to access the significance of the difference between using
$V_{\rm c}(\langle v^2\rangle)$ and $V_{\rm c}(M,z)$, we plotted
$V_{\rm c}(M,z)$ versus $V_{\rm c}(\langle v^2\rangle)$ (Fig.\ A1).
Clearly, on {\it average} the spherical infall model is a good
approximation. The scatter represents the spread in half-mass radii
and velocity dispersions for a given mass found in the simulations.
An upper limit to the mass of a galaxy halo particle needs to be set,
as discussed in detail in Section 3.3.4.



In practice, we set the upper mass limit to be
$7\times 10^{13} h^{-1}$ M$_\odot$, or about 1600 of the initial
simulation particles. Although justifyable in numerical terms,
this mass is also close to the one that results from the cooling
criterion $t_{\rm cool} < t_{\rm coll}$, where $t_{\rm coll}$ is
the halo collapse time for the spherical collapse model
(Rees \& Ostriker 1977; Binney 1977; Silk 1977).
For primordial abundances, this limit is about
$1.5\times10^{13} \Omega_{\rm b} h^{-1}$M$_\odot$ (White 1996),
i.e.\ $\approx 1.0 h^{-1}\times10^{12}$M$_\odot$ for our choice
of $\Omega_{\rm b}$.
For enriched gas, with abundances near solar, the limit goes up to
$5-10\times 10^{13} h^{-1}$M$_\odot$.

\section*{\it Formation and merging of galaxies within dark haloes}

\tx For the modelling of the galaxies that populate the dark matter haloes
we follow the approach of CAFNZ. In their scheme, galaxies
form in the centre of dark haloes. When dark haloes merge, their galaxies
merge with a delay giving by the dynamical friction timescale.
During this dynamical friction phase a dark halo can thus contain several
galaxies at the same time. One of these galaxies will be designated as the
galaxy to which halo hot gas can cool, while the others are considered
satelites. It is important to realize that this part of the CAFNZ
prescription is kept identical, whereas the formation and evolution
of the dark halo population is different.
Thus, dynamical friction is assumed to delay the merging of galaxies
within merged haloes, with the delay given by
$$\tau_{\rm mrg}=\tau^0_{\rm mrg}(M_{\rm halo}/M_{\rm gal})^{\alpha_{\rm mrg}}
  \ . \eqno(A\neweq)$$
In the following, all gas dynamics and star formation happens in between
`major events', which are either the formation of a halo (plus galaxy),
the merger of one or more haloes, or the merger of one or more galaxies
within an already merged halo. These three events define lifetimes for
haloes and galaxies over which gas can cool, stars can form, etc.,
according to processes we describe next.

\section*{\it Stars and gas}

\tx Gas dynamics and star formation are modelled using analytical
relations that are power-law functions of the halo circular
velocity, $V_{\rm c}$, which is a constant for a given halo as
we assume an isothermal density profile.

Each galaxy has a reservoir of stars, cold gas, and hot gas.
A newly formed galaxy has all its baryons, with a total mass of
$\Omega_b M_{\rm halo}$, in the form of hot gas. The gas temperature
is assumed to quickly settle to the virial temperature, which is a
function of $V_{\rm c}$ only:
$$ T_{\rm gas} = T_{\rm vir} = 
  {\mu m_{\rm p}\over 2 k} V_{\rm c}^2. \ \eqno(A\neweq)$$
The three constants $\mu$, $m_{\rm p}$ and $k$ are the mean
molecular weight, proton mass, and Boltzmann's constant, respectively.
Hot gas will cool radiatively at a rate which 
depends on the density $\rho$, temperature $T$, and
metallicity $Z$ of the hot gas. The cooling time-scale is given by
$$\tau_{\rm cool}={3\over 2}{\rho_{\rm gas}(r)\over \mu m_{\rm p}}
  {k T_{\rm gas}\over n_{\rm e}^2(r)\Lambda(T_{\rm gas},Z_{\rm gas})},
  \ \eqno(A\neweq)$$
where $n_{\rm e}$ is the electron density, and $\Lambda(T,Z)$ the
cooling function. For the latter we take the cooling functions for
primordial and solar metallicities as given by Sutherland \& Dopita (1993),
and obtain the cooling function for any given metallicity by linear
interpolation or extrapolation of $\log \Lambda(T,0.0002)$ and
$\log \Lambda(T,Z_\odot)$.

For gas at a constant temperature with a homogeneous chemical abundance,
the cooling rate depends only on the density, and is therefore a function
of radius. Thus, at a large enough radius $r_{\rm cool}$ the cooling
time will be larger than the time passed between two major halo events
(formation or merger).
The mass encompassed by this cooling radius is transferred from the hot
to the cold gas reservoir between these two events.

Stars can form from cold gas, thus transferring mass from the cold
gas reservoir to the stellar population of the galaxy. In the present
modelling, mass ejected by dying stars is {\it not}\ transferred back
to any of the gas reservoirs. However, cold gas can be reheated and thus
transferred back to the hot gas reservoir by the energy output of
supernovae. This process is called `feedback' for short.

The star formation rate is proportional to the amount of cold gas
actually available:
$$d M_*(t,V_{\rm c}) / d t = M_{\rm cold}(t,V_{\rm c}) / \tau_*(V_{\rm c})
  \ , \eqno(A\neweq)$$
with the star formation time-scale given by
$$\tau_*(V_{\rm c})=\tau_*^0 (V_{\rm c}/ 300\ {\rm km\ s}^{-1})^{\alpha_*}
  \ .  \eqno(A\neweq)$$
It is assumed that the reheating of cold gas by supernovae can be modelled as
$$d M_{\rm hot}(t,V_{\rm c}) / d t = \beta(V_{\rm c}) d M_*(t,V_{\rm c}) / d t
  \ , \eqno(A\neweq)$$
where the feedback proportionality parameter $\beta$ is given by
$$\beta(V_{\rm c})=(V_{\rm c}/V_{\rm hot})^{-\alpha_{\rm hot}}
 \ . \eqno(A\neweq)$$
The overall feedback parameter $f_v$, defined as the fraction of energy
released by supernovae that is dumped into the cold gas reservoir as
kinetic energy, determines the values of the constants
$\alpha_*$, $\alpha_{\rm hot}$, and $V_{\rm hot}$.
In this paper we mostly use $f_v=0.2$, for which CAFNZ found the following
values from fits to simulation data by Navarro \& White (1993):
$\alpha_*=-1.5$, $\alpha_{\rm hot}=5.5$, and $V_{\rm hot}=140$ km s$^{-1}$.
For a detailed description of the feedback process and its parameters
we refer to CAFNZ.

The merging of haloes and galaxies also determine where stars, cold,
and hot gas end up. A merger of haloes brings together all amounts of hot
gas of the merging haloes, and reheat the new single hot gas reservoir to
the virial temperature of the newly formed halo. It also brings together
all galaxies that were part of the original haloes. The most massive of these
will now be the one to which hot gas can cool and form stars. CAFNZ set
the circular velocity of this halo to be that of the newly formed dark
halo, which can be rather large, unless its merger time
$\tau_{\rm mrg}$, as given by eq.\ (A4), is larger than the time available
until the next merger event (or the present epoch), in which case the
original $V_{\rm c}$ is retained.

All other galaxies retain their identity (their stars and cold gas
reservoir) if they do not merge with the most massive one, i.e.\ if
$\tau_{\rm mrg}$ is smaller than the time available until the next
merger event (or the present epoch).
In that case stars keep forming from the cold gas left in each of
those galaxies, whereas feedback not only reheats part of the gas,
but also expells it to the hot gas reservoir of the common dark halo.

\section*{\it Stellar population models}

\tx To predict the spectra and photometric properties of galaxies we have used
the extensive library of synthetic stellar population models computed by
Jimenez et al.\ (1998; 1999), where a detailed description of the models can
be found. Here we briefly review the main ingredients of the models. We
computed simple synthetic stellar populations (SSPs), i.e. stellar populations
with fixed metallicity and formed in a burst of infinitesimal duration, for
metallicities between 1/100 Z$_{\odot}$ and 2 Z$_{\odot}$ and ages between 0.01
and 14 Gyr.  The library uses new interior models (Jimenez et al.\ 1998) and a
new set of stellar photospheres for the coolest models (T$_{\rm eff} < 6000$),
while Kurucz (1992) models are used for T$_{\rm eff} > 6000 $K.
Two important new features of the models
are the inclusion of an accurate treatment of all post-main sequence
evolutionary stages and the use of better theoretical photospheric models
for low temperatures.

Simple stellar populations (SSP's) are the building blocks of any
arbitrarily complicated population since
the latter can be computed as a sum of SSP's, once the star formation rate is
provided. In other words, the luminosity of a stellar population of age $t_0$
(since the beginning of star formation) can be written as:
$$L_{\lambda}(t_0)=\int_{0}^{t_0} \int_{Z_i}^{Z_f}
  L^{\rm SSP}_{\lambda}(Z,t_0-t)\, dZ\, dt \ \eqno(A\neweq)$$
where the luminosity of the SSP is:
$$L^{\rm SSP}_{\lambda}(Z,t_0-t)= \int_{m_{\rm d}}^{m_{\rm u}} \dot M_*(Z,m,t)\,
l_{\lambda}(Z,m,t_0-t)\, dm\ \eqno(A\neweq)$$ and $l_{\lambda}(Z,m,t_0-t)$ is
the luminosity of a star of mass $m$, metallicity $Z$ and age $t_0-t$, $Z_i$
and $Z_f$ are the initial and final metallicities, $m_{\rm d}$ and $m_{\rm u}$
are the smallest and largest stellar mass in the population and
$d M_*(Z,m,t) / d t$ is the star formation rate at the time $t$ when the
SSP is formed.

We assume that the initial mass function (IMF) is universal and is given by
a Salpeter IMF ($x=1.35$). In order to take account for the chemical evolution
we have computed detailed models that link $d M_* / d t$ and $Z$, i.e.\
$M_*(Z,t)$ since we have assumed that the $m$ dependence is constant
(a constant IMF). This is described in Appendix C.

\section*{Appendix B: starbursts}

\eqnumber=1
\tx Besides a quiescent mode of star formation, we consider a bursting
mode of star formation associated with a major formation or merging event.
This star formation mode can be the dominant one, especially at early times
when encounters and mergers are frequent, and most galaxies form.
The strength of the starburst will depend on the clumpiness
of the matter distribution that is induced by an encounter or merging
event, on the duration of the starburst phase, and on the amount of cold
gas available to form stars, which can be substantially enhanced by strong
inflows of gas into the central region of the galaxy during the final stages
of a merger event (Mihos \& Hernquist 1996).
Here we adopt a simplified approach, by assuming that the SFR during
the starburst phase can be modelled using the same scaling relations
as for the quiescent star formation mode, but with different
star formation and cooling rates. The feedback strength is assumed to be the
same for both star formation modes. Also, we adopt the same Salpeter
IMF for both modes, a choice that is supported by recent observational
data (see Elmegreen 1999).

The duration of the starburst, $t_{\rm b}$, is assumed to be equal to
the dynamical timescale of the new dark matter halo:
$t_{\rm b}=\tau_{\rm dyn}\equiv R_{\rm h}/\sigma_{\rm v}$. Using the
spherical infall model (eq.\ A1), the virial theorem (eq.\ \the\virialB),
and the isothermal profile (eq.\ A2), we find
\newcount\bursttime
\bursttime=\eqnumber
$$t_{\rm b}\approx 0.024 (1+z_{\rm form})^{-3/2} h^{-1}{\rm Gyr}\ .
  \eqno(B\neweq)$$
Therefore, a single starburst at low redshift lasts longer than one
at high redshift, but as the frequency is much lower, it will still
be at high redshifts that starbursts are more important.
Note that the actual starburst might last for a much shorter time
than the dynamical timescale, depending on the amount of material
available in the bulges of the merging galaxies (Mihos \& Hernquist 1994),
so $t_{\rm b}$ should be taken as a maximum.

The bursting star formation timescale $\tau_{*,\rm b}$ is assumed to
scale with the halo circular velocity $V_{\rm c}$ in the same manner as
the continuous star formation timescale $\tau_*$, i.e.\
$$\tau_{*,\rm b}(V_{\rm c})= \tau^0_{*,\rm b}
  (V_{\rm c}/ 300\ {\rm km\ s}^{-1})^{\alpha_*}\eqno(B\neweq)$$
(analogous to eq.\ (A8) in Appendix A).

It is also necessary to introduce a `burst factor' $f_{\rm b}$ which
models the enhanced
cooling of hot gas and stronger inflow of cold gas during the
star bursting phase. It is expressed as a density enhancement over and
above the mean isothermal profile that will form after the bursting event.
If $f_{\rm b}$ is larger than unity, cooling will be more efficient,
so that star formation will be enhanced during the bursting phase.

Thus, during a starburst of duration $t_{\rm b}$, star formation is
enhanced by having more cold gas available to form stars, and by having
those stars formed at a higher rate. The shorter star formation timescale
will result in a different stellar population as compared to the continuous
mode for two reasons: firstly and simply due to the different formation
timescale, but secondly because the chemical evolution of the population
will be different.


\figps{6}{S}{\psfig{file=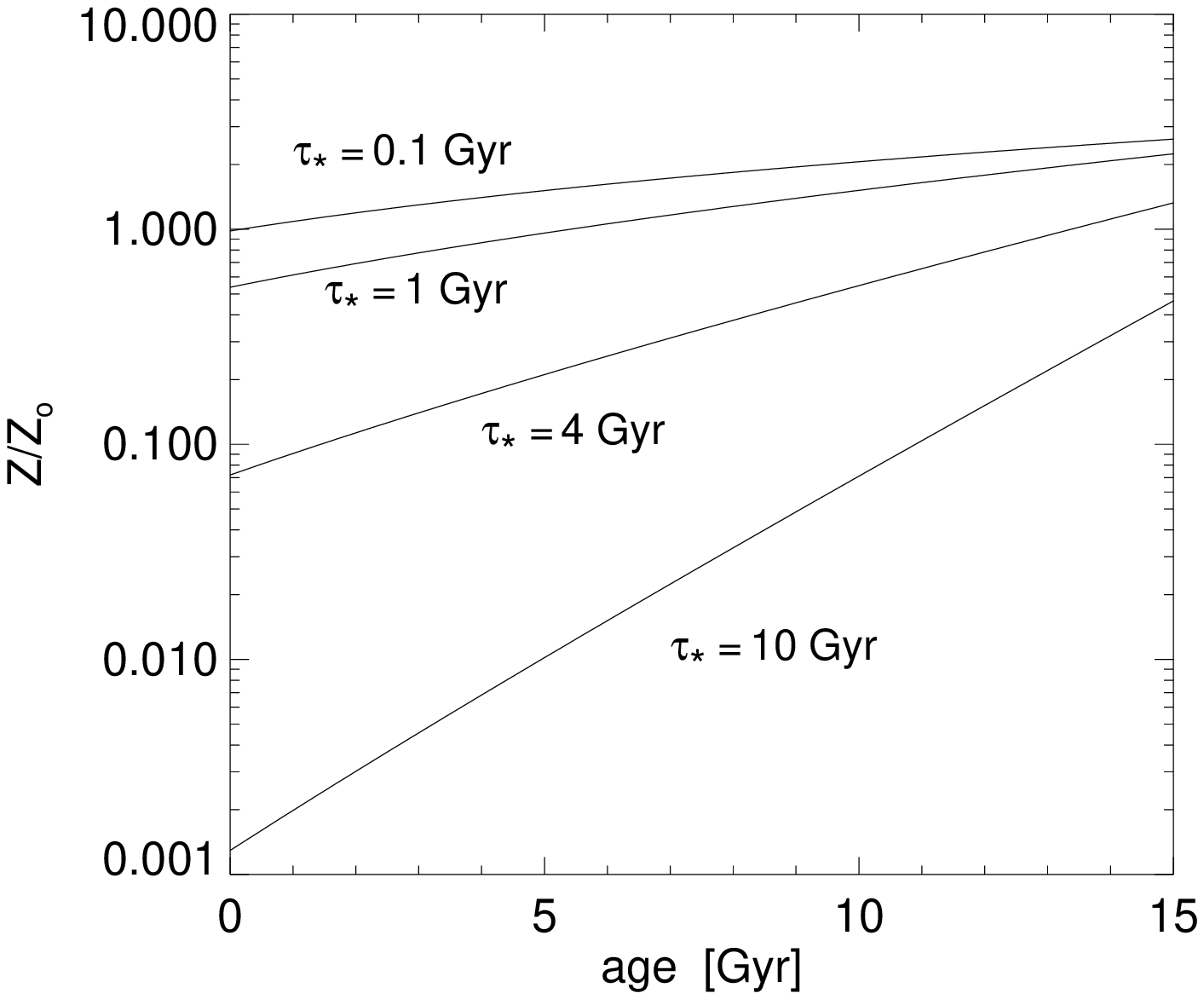,width=8.5cm,silent=}}
{{\bf Figure C1.} Metallicity evolution for various star formation
time-scales $\tau_*$.}

\figps{7}{S}{\psfig{file=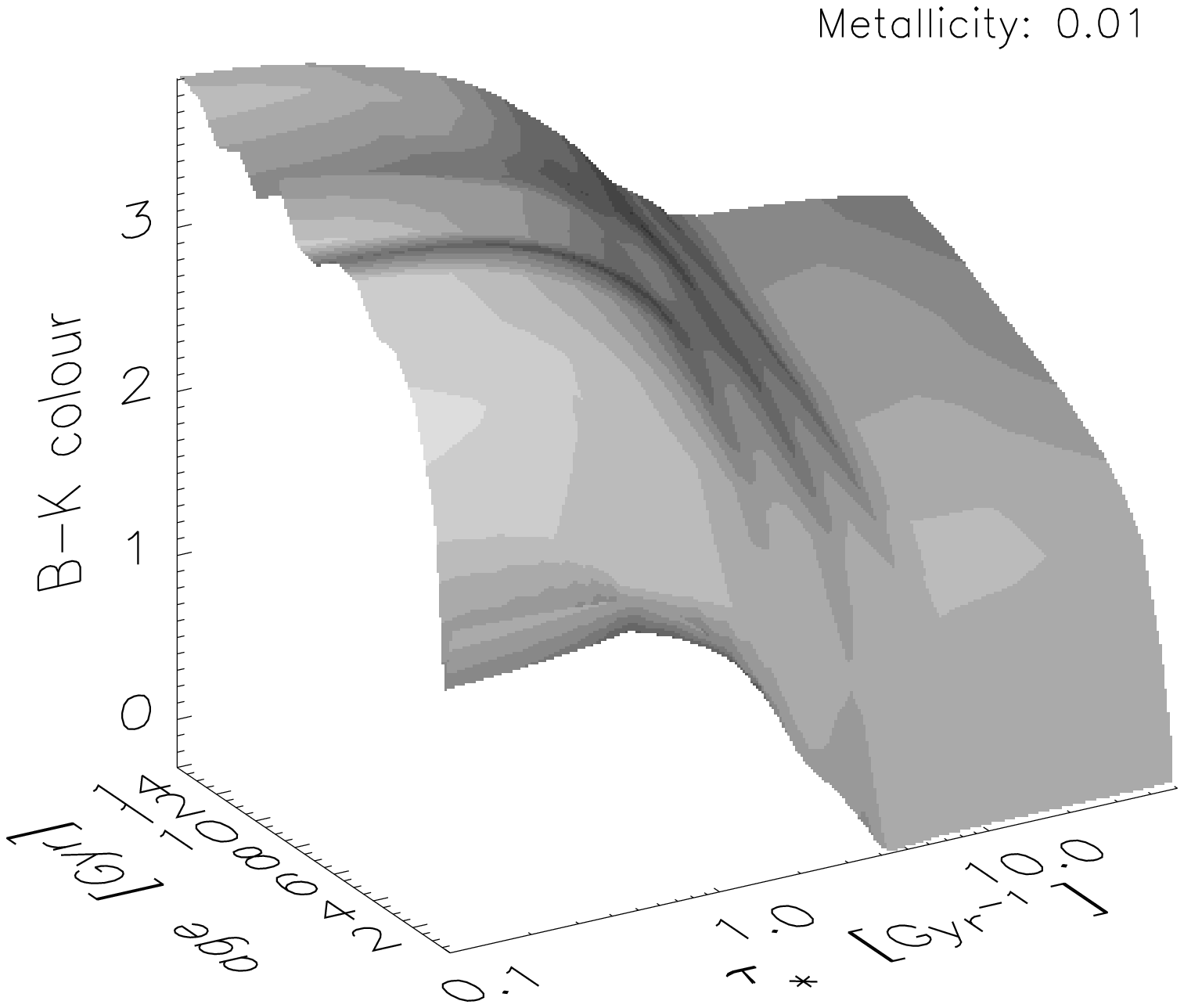,width=9.0cm,silent=}}
{{\bf Figure C2.} B-K colour for a stellar population with an initial
metallicity of 0.01, as a function of age and star formation
time-scale $\tau_*$.}

\section*{Appendix C: chemical evolution}

\eqnumber=1

\tx Most of the phenomenological models published so far make the
simplifiying assumption that hot gas cools with primordial metallicity,
while the stellar populations form and evolve at a constant solar
metallicity. This `instant' enrichment scheme helps to produce galaxies
that are red enough in order to match observations, but is only realistic
for star bursting galaxies, not for galaxies that form stars more
continuously. Furthermore, the hot gas around metal-ejecting galaxies
is enriched, which will boost cooling at intermediate and low redshifts.
We therefore include chemical enrichment in our modelling,
in the following way.

The chemical evolution of a stellar population can be divided into two
parts: the chemical enrichment rate at which the gas is polluted
by dying stars and the formation rate at which gas is transformed into stars.
In reality these two aspects are related, since the more metals are injected
into the interstellar medium, the more efficient star formation should be,
and also vice versa (e.g.\ low surface brightness galaxies have low
metallicity and low star formation rates). In this paper, we assume that
the star formation rate depends solely on the circular velocity of the halo
(see Appendix A).
We then compute the enrichment of the interstellar medium using accurate
nucleosynthesis prescriptions. We refer to Jimenez et al.\ (1999) for a
detailed description of the chemical evolution equations used and the stellar
yields.

In principle, one should compute the hydrodynamical evolution of the baryons
for each halo and calculate the surface density of the baryonic disc in order
to determine the star formation rate as well as the infall rate and the
outflow from supernovae. Since our simulations do not have a hydrodynamical
description of the baryonic component, we proceed in a different way.
We use the chemical evolution models of Matteucci \& Francois (1989),
but with a star formation rate given by eq.\ (A7), thus implicitely
assuming a specific disk model. As our star formation rate is similar
to that of e.g. Heavens \& Jimenez (1999) and Jimenez et al.\ (1998),
this disk model will be close to the disk model of these authors.
With this set-up it is possible
to obtain a surface that describes $\tau_*$ vs. $Z$(t), where $\tau_*$
is given in terms of $V_{\rm c}$ by eq.\ (A8). We show $Z$(t) for four
different values of $\tau_*$ in Fig.\ C1. For a given $\tau_*$ and initial $Z$,
a population will evolve with time as described in Appendix A.
An example of the B-K colour evolution is shown in Fig.\ C2.


\section*{Appendix D: star formation below the N-body resolution limit}

\tx The N-body simulation technique has a lower limit on the mass of the
haloes identified. In the hierarchical picture these must have formed
through merging of haloes with masses below that limit. We thus need to
approximate the stellar and gas mass resulting from the evolution along
the merger tree below the numerical resolution limit.
Both modes of star formation need to be accomodated, with the complication
that merging of small-scale structure is relatively frequent at early
times. Although both modes will occur intermittently, we approximate the
average of all these periodic contributions taken together, by assuming
a basic star formation timescale of 0.5 Gyr.
We estimate the amount of gas cooled in the sub-tree by considering
a single population which has two thirds of the circular velocity of
the halo in which the sub-tree terminates, and half its age at the
time of formation.

This approximation will be improved upon in future work
by linking analytical merger subtrees, obtained using the extended
Press-Schechter formalism, to the leaves of the merger tree
extracted from the numerical simulation.
For the present paper, these subtrees below the N-body resolution
limit are not important, as they represent old stellar populations
that have faded over many Gyr, and are likely to end up in bulges and
centres of elliptical galaxies. They therefore contribute little
to the present day luminosities that enter the Tully-Fisher relation
and luminosity functions.


\end